\documentclass[]{aa}
\usepackage{graphicx}
\usepackage{natbib}
\def\ebv{\mbox{$E_{B-V}$}}

\def\lya{\mbox{Ly$\alpha$}}

\def\ecs{\mbox{~erg~cm$^{-2}$~s$^{-1}$}}

\def\lesssim{\mathrel{\hbox{\rlap{\hbox{\lower4pt\hbox{$\sim$}}}\hbox{$<$}}}}
\def\gtrsim{\mathrel{\hbox{\rlap{\hbox{\lower4pt\hbox{$\sim$}}}\hbox{$>$}}}}
\def\nhi{\mbox{$N(\ion{H}{i})$}}

\usepackage{txfonts}
\begin{document}
\title{An integral field spectroscopic survey for high redshift damped Lyman-$\alpha$ galaxies
 \thanks{Based on observations collected at the Centro Astron\'omico
   Hispano Alem\'an (CAHA), operated by the Max-Planck Institut f\"ur
   Astronomie and the Instituto Astrofisica de Andalucia (CSIC).}
}
\titlerunning{An IFS survey for high-$z$ DLA galaxies}
   \author{L. Christensen          \inst{1}
           \and L. Wisotzki         \inst{2}
           \and M.~M. Roth          \inst{2}
          \and S.~F. S\'anchez     \inst{3}
          \and A. Kelz    \inst{2}
          \and K. Jahnke   \inst{4}
          }

   \offprints{L. Christensen}  
   \institute{European Southern Observatory, Casilla 19001, Santiago 19, Chile,
     \email{lichrist@eso.org}
   \and  Astrophysikalisches Institut Potsdam, An der Sternwarte 16,
     14482 Potsdam, Germany
   \and  Centro Astron\'omico Hispano Alem\'an de Calar Alto, Calle Jes\'us Durb\'an Rem\'on 2,2 E-04004 Almer\'ia, Spain
    \and  Max-Planck-Institut f\"ur Astronomie, K\"onigstuhl 17, 69117
  Heidelberg, Germany  }
   \date{Received/accepted}
   
     
   \abstract
{}
{We search for galaxy counterparts to damped Lyman-$\alpha$ absorbers
  (DLAs) at $z>2$ towards nine quasars, which have 14 DLAs and 8
  sub-DLAs in their spectra.}
{We use integral field spectroscopy to search for \lya\ emission line
  objects at the redshifts of the absorption systems.}
{Besides recovering two previously confirmed objects, we find six
  statistically significant candidate \lya\ emission line objects. The
  candidates are identified as having wavelengths close to the DLA
  line where the background quasar emission is absorbed. In comparison
  with the six currently known \lya\ emitting DLA galaxies the
  candidates have similar line fluxes and line widths, while velocity
  offsets between the emission lines and systemic DLA redshifts are
  larger.  The impact parameters are larger than 10 kpc, and lower
  column density systems are found at larger impact parameters.}
{ Assuming that a single gas cloud extends from the QSO line of sight
  to the location of the candidate emission line, we find that the
  average candidate DLA galaxy is surrounded by neutral gas with an
  exponential scale length of $\sim$5~kpc.}


\maketitle
%

%

\section{Introduction}
Galaxy counterparts to Damped Lyman-$\alpha$ systems (DLAs) seen in
quasar (QSO) spectra have been suggested to be (proto)-disk galaxies
with line of sight clouds of neutral gas with column densities
\nhi~$>2\times10^{20}$ cm$^{-2}$ \citep{wolfe86}.  Analyses of
absorption line profiles have indicated that rotational components
with velocities of $\sim$200~km~$^{-1}$ can be involved in these
systems suggesting that DLAs reside in large disk galaxies
\citep{prochaska97,ledoux98}. On the other hand, numerical simulations
show that in a hierarchical formation scenario merging proto-galactic
clumps can also give rise to the observed line profiles
\citep{haehnelt98}.

A large fraction of the neutral hydrogen present at $z>2$ is contained
in high column density DLA systems
\citep{lanzetta95,storrie00,peroux01}.  In addition to the classical
DLAs, clouds with column densities
$10^{19}~<~\textrm{\nhi}~<~2\times~10^{20}$~cm$^{-2}$ also show some
degree of damping wings, which is characteristic of DLA systems. It is
suggested that sub-DLA systems contain a significant fraction of the
neutral matter in the Universe \citep{peroux03}. Metallicity studies
have shown that the properties of the sub-DLA systems are similar to
those of DLA systems \citep{dessauges03}, apart from the latter
category having large ionisation corrections \citep{prochaska04}.

The association of DLAs with galaxies has been a subject of much
study.  Originally, either space-based or ground-based deep images
were obtained to identify objects near the line of sight to the QSOs
\citep{steidel95,lebrun97,warren01}. To confirm nearby objects as
galaxies that are responsible for the DLA lines in the QSO spectra,
follow-up spectra are needed to find the galaxy redshifts. At $z<1$,
confirmations of 14 systems exist to date \citep[][and references
  therein]{rao03,chen03,lacy03,chen05}, while at $z\gtrsim2$ only 6
DLA galaxies are confirmed through spectroscopic observations of
\lya\ emission from the DLA galaxies
\citep{moller93,djorgovski96,moller98,leibundgut99,moller02,moller04},
three of which are located at the same redshifts as the QSOs
themselves. Other techniques to identify DLA galaxies involve
narrow-band imaging \citep[e.g.][]{fynbo99,fynbo00} or Fabry-Perot
imaging.  A Fabry-Perot imaging study of several QSO fields did not
result in detections of emission from DLA galaxies
\citep{lowenthal95}, while recently the same method was used to
identify a few emission line candidates \citep{kulkarni06}.

Integral field spectroscopy (IFS) presents an alternative that
provides images and spectra at each point on the sky
simultaneously. This technique can be used to look for emission line
objects at known wavelengths, but unknown spatial location. This
technique is ideally suited to look for \lya\ emission lines from the
galaxies responsible for strong QSO absorption lines.  At the
\lya\ wavelength corresponding to the redshift of the DLA system, the
QSO emission has been absorbed, enabling us to locate emission line
objects very near to the QSO line of sight.  Because of the large
column densities in DLAs and the resonant nature of \lya\ photons the
corresponding emission line may be offset in velocity space relative
to the DLA line \citep[e.g.][]{leibundgut99}, but such an offset is
not always observed \citep[e.g.][]{moller04}.

IFS searches for emission from DLA galaxies towards two QSOs have
resulted in upper limits for their fluxes
\citep{petitjean96,ledoux98b}, while a sub-DLA galaxy previously known
to be a \lya\ emitter was confirmed with IFS \citep{christensen04a}.
Here we present a survey using IFS to look for \lya\ emitting DLA
galaxies. Section~\ref{sect:hiz_sample} describes the sample of QSOs
included in the survey, which are known previously to have DLAs and
sub-DLAs in their spectra. Section~\ref{sect:hiz_obs} describes the
observations and data reduction. In Section~\ref{sect:hiz_cand} the
method of detecting emission line candidates is described.
Section~\ref{sect:hiz_results} presents the results and comments on
each object. Properties of the \lya\ emission candidates detected in
the survey in relation to the six previously known \lya\ emitting DLA
galaxies are presented in Section~\ref{sect:hiz_proper}.
Section~\ref{sect:hiz_conc} summarises our findings.  A flat cosmology
with $H_0=70$ km~s$^{-1}$~Mpc$^{-1}$, $\Omega_m=0.3$, and
$\Omega_{\Lambda}=0.7$ is used throughout.

This study, as well as previous ones that try to identify the host
galaxies of DLA systems, can be biased since the galaxy observed at
the right redshift likely belongs to the brightest emission line object
close to the line of sight. In the case that the host galaxy is a much
fainter galaxy in a group, it will not be identified correctly.  In
the remaining part of the paper, an `identified' DLA galaxy refers to
observations that show (line) emission from independent observations,
while the `candidates' are only reported in these IFS
observations. Although extensive tests are done on the data to
distinguish the candidates from potential artifacts, independent
observations are needed to prove them as \lya\ emitters connected with
the DLAs.

\section{Sample selection}
\label{sect:hiz_sample}

We selected a number of DLA systems without previous detections of
associated \lya\ emission. The selected QSOs with known DLAs were
chosen based on the following criteria
\begin{enumerate}
\item \nhi $> 2\times10^{20}$ cm$^{-2}$
\item DLA redshift ($2<z<4$) 
\item Northern hemisphere object 
\end{enumerate}

The first criterion includes only classic DLA systems. Many QSO
spectra show additional sub-DLA systems. Although their relationship
to DLAs is still debated, these systems were included in the survey
because of their probable physical association with galaxies.

To increase the sample size with a minimum number of pointings we
preferentially selected QSOs with multiple DLAs. IFS covers a range of
wavelengths, and correspondingly \lya\ emission at a large range of
redshifts in the line of sight for each QSO. However, in retrospect,
this can affect the emission line detections, because extinction in
foreground DLAs could obscure emission from background ones when the
galaxies lie in the same line of sight. Hence, upper limits on
detections of the higher redshift systems can be biased.

From the list of DLA systems compiled by S.~Curran\footnote{\tt
  http://www.phys.unsw.edu.au/$\sim$sjc/dla/}, we found 66 QSOs
matching these criteria in 2003.  More recently, detections of DLAs in
the Sloan Digitized Sky Survey QSO spectra have greatly increased the
number of known DLAs \citep{prochaska04,prochaska05}. A systematic
survey of all 66 objects would require a large amount of time with
present instruments, so we selected a few systems based on their
observability during the allocated observing runs.  We avoided DLAs
with \lya\ absorption lines close to sky emission lines.

The total sample consists of 9 QSOs with a total number of 14 DLA
systems as listed in Table~\ref{tab:list_obj}.  These QSOs have an
additional 8 sub-DLAs which are included in the survey. Because of the
small number of DLAs involved in the survey, a proper statistical
study is not the aim of this paper.  Instead we focus on a few systems
to exploit the benefits of IFS for this kind of investigation.

To study the applicability of IFS in identifying DLA galaxies we
initially observed DLA galaxies where \lya\ emission had been reported
previously in the literature.  Two of these systems could be observed
during our runs; {Q2233+131} and {PHL 1222}, originally
identified by \citet{steidel95}, \citet{djorgovski96} and
\citet{moller98}. Both objects are reported to have extended
\lya\ emission
\citep{fynbo99,christensen04a}. Table~\ref{tab:list_obj} includes
these two previously known \lya\ emitting DLA and sub-DLA galaxies,
although the criteria listed above are not satisfied. The absorption
system towards Q2233+131 has a column density that classifies it as a
sub-DLA.  Unless otherwise noted, these two objects are kept separate
from the detection of candidate emission line objects in the remainder
of the paper.
  
Most of the DLAs in the IFS study lie towards bright QSOs ($R<19$).
This ensured that the PSF variations as a function of wavelength could
be determined, which was necessary for the subtraction of the QSO
emission.  Bright QSOs had larger residuals from the subtraction of
the continuum emission, which potentially affected out ability to
recover emission line objects that were offset in velocity space and
located closer than 1\arcsec\ to the QSO line of sight. However, tests
with artificial objects showed that this was a minor problem for the
data set (see Sect.~\ref{sect:hiz_art}).

\begin{table*}
\centering
\begin{tabular}{llllllll}
\hline
\hline
  \noalign{\smallskip}
 Coordinate name & Alt. name   & $z_{\mathrm{em}}$ &
 $z_{\mathrm{abs}}$ & $\log$\nhi &   [Fe/H] & [Si/H]  & Ref.\\
  \noalign{\smallskip}
\hline
  \noalign{\smallskip}
Q0151+048A &{PHL~1222}   & 1.93 & 1.934 & 20.36$\pm$0.10 & & & (1)\\
   \noalign{\smallskip}
Q0953+4749 & {PC 0953+4749} & 4.457& 3.404 & 21.15$\pm$0.15 &
   $>$--2.178 & $>$--2.09   & (2,3)\\
                                   &  &   & 3.891 & 21.20$\pm$0.10 &
  $>$--1.712  & $>$--1.60 \\
  &                    &                  & 4.244 & 20.90$\pm$0.15 &
   --2.50$\pm$0.17 &--2.23$\pm$0.15 \\
   \noalign{\smallskip}
{Q1347+112} &  & 2.679&  2.471 & 20.3     & &  &  (4,5,6)\\
   &                &      &  2.05  & 20.3$^\dagger$ & &  &  (7)\\
   \noalign{\smallskip}
Q1425+606& {SBS 1425+606}  &  3.163 & 2.827 & 20.30$\pm$0.04
   & --1.33$\pm$0.04 & $>$--1.03 & (8,9,10,21) \\
   \noalign{\smallskip}
{Q1451+1223}& B1451+123 & 3.246 &  2.469 & 20.39$\pm$0.10 &
   --2.54$\pm$0.12 & --1.95$\pm$0.16  & (11,24)\\
  &  &  &  3.171 & 19.70$\pm$0.15 & --1.87$\pm$0.16 & --1.62$\pm$0.15 & (19) \\
   & &  &  2.254 & 20.30$\pm$0.15 & --1.47$\pm$0.17 & $>$--0.40 &  (6,12,19)\\
   \noalign{\smallskip}
{Q1759+7539} & {GB2 1759+756}  &  3.05 &    2.625 & 20.76$\pm$0.10
   & --1.21$\pm$0.10 &--0.82$\pm$0.10 & (13,18,20)\\
  &          &    & 2.91 &  19.8 & --1.65$\pm$0.01 & --1.26$\pm$0.01 & (18)\\
   \noalign{\smallskip}
Q1802+5616 & {PSS J1802+5616} & 4.158 & 3.391 & 20.30$\pm$0.10 &
 --1.54$\pm$0.11& $>$--1.55 & (23)\\ 
  &  &  &   3.554 & 20.50$\pm$0.10 &  --1.93$\pm$0.12& $>$--1.82\\
  &  &  &   3.762 & 20.55$\pm$0.15 &  --1.82$\pm$0.26& $>$--1.74\\
  &  &  &   3.811 & 20.35$\pm$0.20 &  --2.19$\pm$0.23& --2.04$\pm$0.22\\
   \noalign{\smallskip}
Q2155+1358 & {PSS J2155+1358} & 4.256 & 3.316 & 20.55$\pm$0.15
& $>$--1.68 & --1.26$\pm$0.17 & (3)\\
  &  &  &   3.142 & 19.94$\pm$0.10 & --2.21$\pm$0.21 & --1.85$\pm$0.12 & (14,19) \\
 &   &  &   3.565 & 19.37$\pm$0.15 & $<$--2.40 & --1.27$\pm$0.16 & (14,19)\\
 &   &  &   4.212 & 19.61$\pm$0.15 & --2.18$\pm$0.25 & --1.92$\pm$0.11 & (14,19)\\
   \noalign{\smallskip}
{Q2233+131}&  & 3.295 & 3.153 & 20.0   & $>$--1.4$\pm$0.1 & $>$--1.04 &(15,16,17) \\
 &   &  & 2.551 & 20.0   & &  & (12,16)\\
   \noalign{\smallskip}
\hline
\end{tabular}
\caption{List of the observed DLA and sub-DLA systems with column
  densities and metallicities taken from the literature. $^\dagger$
  denotes a system where the reported \nhi\ needs to be confirmed
  through high resolution spectroscopy. References for either DLA
  redshifts or metallicities: (1) \citet{moller98}, (2)
  \citet{schneider91}, (3) \citet{prochaska03b}, (4) \citet{smith86},
  (5) \citet{wolfe86}, (6) \citet{turnshek89}, (7) \citet{wolfe95},
  (8) \citet{chaffee94}, (9) \citet{stephanian96}, (10)
  \citet{prochaska02}, (11) \citet{bechtold94}, (12)
  \citet{lanzetta91}, (13) \citet{prochaska01}, (14) \citet{peroux03},
  (15) \citet{steidel95}, (16) \citet{lu97}, (17) \citet{lu98}, (18)
  \citet{outram99}, (19) \citet{dessauges03}, (20)
  \citet{prochaska02b}, (21) \citet{lu96}, (22) \citet{prochaska03},
  (23) \citet{prochaska03c}, (24) \citet{petitjean00}.  }
\label{tab:list_obj}
\end{table*}

\section{Observations and data reduction}
\label{sect:hiz_obs}

Using the Potsdam Multi Aperture Spectrophotometer (PMAS) mounted on
the 3.5m telescope at Calar Alto we observed the objects listed in
Table~\ref{tab:hiz_log} during several runs from 2002--2004. The PMAS
integral field unit (IFU) was used in the standard configuration where
256 fibres are coupled to a 16$\times$16 element lens array. During
the observations each fibre covered 0\farcs5$\times$0\farcs5 on the
sky giving a field of view of 8\arcsec$\times$8\arcsec. Each fibre
resulted in a spatial element (spaxel) represented by a single
spectrum. The 256 spectra were recorded on a 2k$\times$4k CCD which
was read out in a 2$\times$2 binned mode. With a separation of 7
pixels between individual spectra, the fibre to fibre cross-talk was
negligible (less than 0.4\% for an extraction of all 7 pixels).
Detailed overviews of the PMAS instrument and capabilities are given
in \citet{pmas00,roth05}.

For individual exposures a maximum time of 1800s was used because of
the large number of pixels affected by cosmic ray hits. Furthermore,
because of varying conditions such as the atmospheric transmission and
seeing, the total exposure time for each object was adjusted, or
sometimes an observation was repeated under better conditions.  The
photometric conditions during observations were monitored in real time
with the PMAS acquisition and guiding camera (A\&G camera) which is
equipped with a 1k$\times$1k CCD. Seeing values listed in
Table~\ref{tab:hiz_log} refer to the seeing \textit{FWHM} measured in
the A\&G camera images.  Determining actual spectrophotometric
conditions requires monitoring of the extinction coefficients which
cannot be determined from the A\&G camera images. In
Table~\ref{tab:hiz_log} `stable' means that the photometry of the
guiding star was constant within 1\% during the observations.

The data were obtained using 2 gratings; one with 300 lines mm$^{-1}$
and one with 600 lines mm$^{-1}$, set at a chosen tilt to cover a
selected wavelength range. The \textit{FWHM} of the sky lines were
measured to be 6.4 and 3.2~{\AA}, respectively.  Observations of
spectrophotometric standard stars were carried out at the beginning
and end of each night at the grating position used for the
observations.

\begin{table*}
\begin{footnotesize}
\centering
  \begin{tabular}{lllllll}
   \hline \hline
   \noalign{\smallskip}
QSO & date & exp.time & grating & $\lambda$ coverage & seeing & conditions\\
    &      & (s)   & & ({\AA})\\
  \noalign{\smallskip}
   \hline
   \noalign{\smallskip}
Q0151+048A     & 2003-08-27 & 5$\times$1800  & V600 & 3500--5080 &0.8--1.2 &
   stable\\
Q0953+4749     & 2004-04-16 & 4$\times$1800  & V300 & 3630--6980 & 0.9 & stable\\
               & 2004-04-21 & 5$\times$1800  & V300 & 3630--6980 & 1.0 & non-phot.\\
Q1347+112      & 2004-04-20 & 7$\times$1800  & V300 & 3630--6980 & 0.6 & non-phot.\\
Q1425+606      & 2004-04-19 & 6$\times$1800  & V300 & 3630--6750 & 1.0 & stable\\
Q1451+1223     & 2004-04-17 & 7$\times$1800  & V300 & 3630--6980 & 0.8 & non-phot.\\
Q1759+7539     & 2004-04-21 & 7$\times$1800  & V300 & 3630--6980 & 1.0--1.5
   & non-phot. \\
Q1802+5616     & 2003-06-18 & 2$\times$1800  & V600 & 5100--6650 & 1.0 & non-phot.\\ 
               & 2003-06-20 & 3$\times$1800  & V600 & 5100--6650 & 1.0 & non-phot. \\
               & 2003-06-21 & 4$\times$1800  & V600 & 5100--6650 & 1.8 & non-phot.\\
               & 2003-06-22 & 6$\times$1800  & V600 & 5100--6650 & 0.9 & stable\\
Q2155+1358     & 2003-08-26 & 7$\times$1800  & V600 & 4015--5610 & 0.7 & stable\\
               & 2003-08-27 & 4$\times$1800  & V600 & 4015--5610 & 0.8 & non-phot.\\
Q2233+131      & 2002-09-03$^\dagger$ & 4$\times$1800  & V300 &
3930--7250 & 1.0--1.3 & \\
               & 2003-08-24 & 6$\times$1800  & V600 & 4000--5600 &  0.6 & stable\\

               & 2003-08-25 & 4$\times$1800  & V600 & 4000--5600 & 0.7 & non-phot.\\
   \noalign{\smallskip}
   \hline
  \end{tabular}
  \caption[]{Log of the observations. The last two columns
   show the average seeing during the integrations and the photometric
   conditions derived from the A\&G camera images. $^\dagger$
   Results from these observations are published in \citet{christensen04a}.}
  \label{tab:hiz_log}
\end{footnotesize}
\end{table*}

Data reduction was done by first subtracting an average bias frame.
Before extracting the 256 spectra most cosmic ray hits were removed by
the routine described in \citet{pych03}. A high threshold was chosen
such that not all cosmic rays were removed, because a low threshold
would also remove bright sky emission lines from some
spectra. Remaining cosmic rays were removed from the extracted spectra
using the program L.A. Cosmic \citep{vandok01}.

The locations of the spectra on the CCD were found from exposures of a
continuum lamp, taken either before or after the science exposures,
using a tracing algorithm developed for the IDL based PMAS data
reduction package P3D \citep{becker02}. The spectral extraction was
done in two ways; a `simple extraction' that added all flux from each
spectrum on the CCD (i.e. an extraction width of 7 pixels), and
another method that took into account the profile of the spectrum on
the CCD. This second method assumed that the spectral profiles were
represented by Gaussian functions (Gaussian extraction) where the
widths were allowed to vary with wavelength. Widths were determined by
fits to each of the 256 spectra as a function of the wavelength, and
the extraction used these width in combination with the centre found
from the tracing algorithm.  The Gaussian profile is an approximation
because the profiles are not strictly Gaussian. The second method
increased the signal-to-noise ratio by $>$10\% for faint objects and
therefore unless otherwise noted, the results from the `Gaussian
extraction' data cubes will be reported \citep[see also][]{sanchez06}.

After extraction, the spectra were wavelength calibrated using
exposures of emission line lamps taken just before or after the
observations. The wavelength calibration was done using the P3D
reduction tool.  Comparisons with sky emission lines indicated an
accuracy of the wavelength calibration of about 10\% of the spectral
resolution.

The spectra show a wavelength dependent fibre to fibre transmission.
To correct for this effect, we extracted sky spectra obtained from
twilight sky observations in the same way as the science
observations. A one dimensional average sky spectrum was calculated.
Each of the 256 spectra were divided by this average spectrum, and the
fraction was fit by a polynomial function to reduce noise. These
polynomials were used to flat field the science spectra.

Before combining individual frames, the extracted spectra were
arranged into data cubes. Each data cube was corrected for extinction
using an average extinction curve for Calar Alto \citep{hopp02}. The
data cube combination took into account a correction for the
differential atmospheric refraction using a theoretical prediction
\citep{filippenko82}. Relative spatial shifts between individual data
cubes were determined from a two-dimensional Gaussian fit to the QSO
PSF at a wavelength close to the strong DLA absorption lines.

Subtraction of the sky background was an iterative process because the
locations of the emission line objects of interest were not known
beforehand. PMAS, in the configuration used, does not have
specifically allocated sky fibres. Instead, we selected 10--20 fibres
uncontaminated by the QSO emission and the average spectrum was
subtracted from all 256 spectra.  Different spaxel selections were
examined visually to select an appropriate sky spectrum which had no
emission line or noisy pixels in the spectral region of interest.

Flux calibration was done in the standard way using observations of
spectrophotometric standard stars. A one-dimensional spectrum of the
standard star was constructed by co-adding flux from all 256
spaxels. This was used to create a sensitivity function that could be
applied to each of the 256 spectra in the science exposures.  For
non-photometric nights the flux calibrated spectra were compared with
QSO spectra from the literature to estimate photometric
errors. However, no correction factor was applied to our spectra,
because an intrinsic variability of the QSOs would make such scaling
uncertain. For some cases we note in Sect.~\ref{sect:hiz_notes} that
there are differences which could be caused by either non-photometric
conditions or intrinsic variability.

\begin{figure*}[!htbp]
\centering
\resizebox{16.5cm}{!}{\includegraphics[bb= 30 60 570 800, clip]{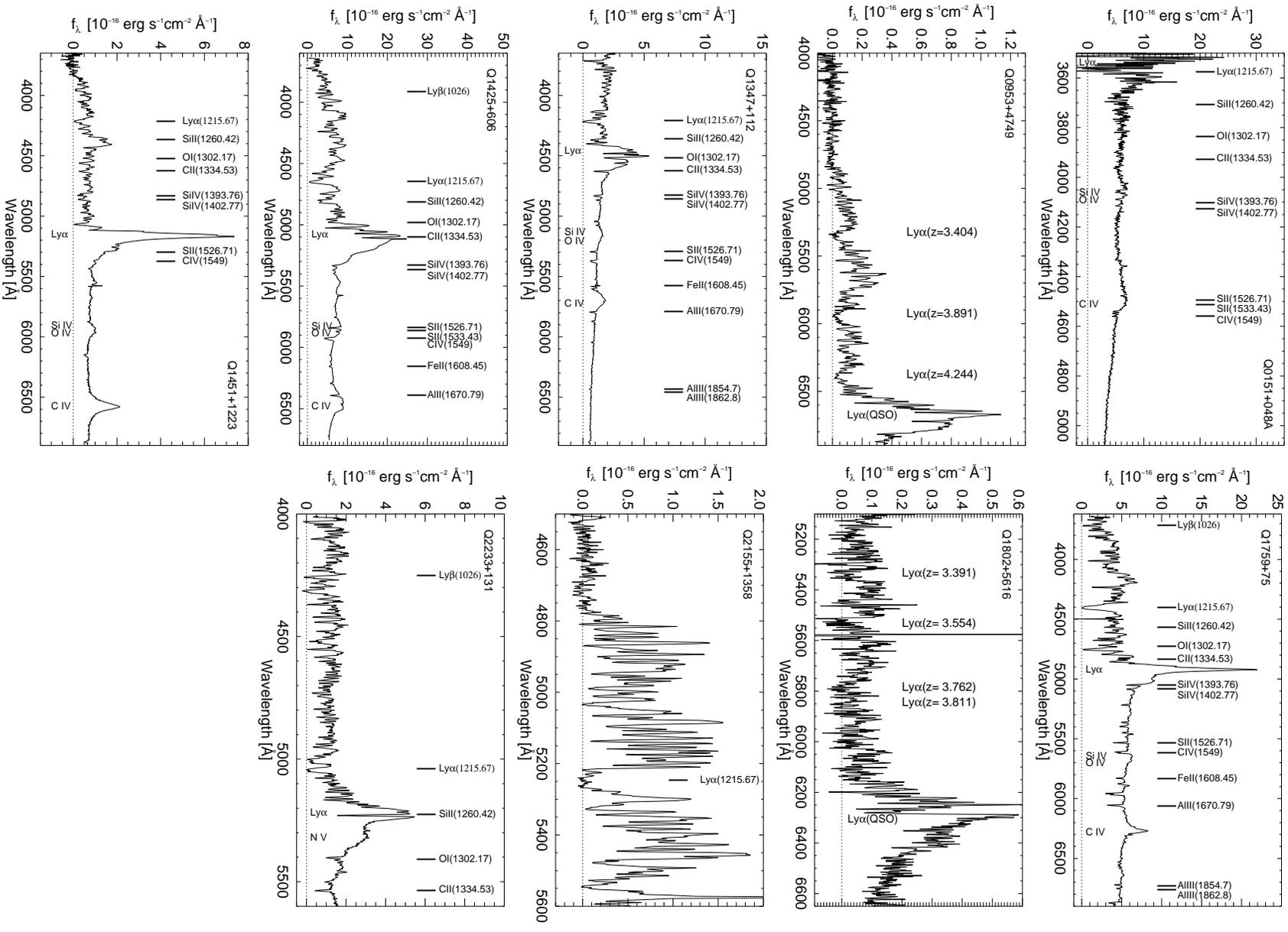}}
\caption{QSO spectra extracted with a 3\arcsec\ radial aperture.  No
  correction for Galactic reddening is applied.  Metal absorption
  lines associated with the DLA systems are indicated. For those QSOs
  with multiple DLA lines only the DLA lines and their redshifts are
  indicated.  Line identifications written below the spectra are QSO
  emission lines. }
\label{fig:hiz_qso_spec}
\end{figure*}

For reference we show spectra of the target QSOs in
Fig.~\ref{fig:hiz_qso_spec}. Where present, metal absorption lines
corresponding to the highest column density DLAs are indicated. For
QSOs with multiple DLAs lines only the DLA lines and their redshifts
are indicated since the wavelength coverage does not include lines
redwards of the QSO \lya\ line. A detailed analysis of metal
absorption lines requires higher resolution spectroscopy as presented
elsewhere \citep[e.g.][]{prochaska03,peroux03,dessauges03}.  DLA
redshifts derived from the metal absorption lines were consistent with
those reported in the literature within the accuracy of the wavelength
calibration of the data cubes.

\section{Search for DLA optical counterparts}
\label{sect:hiz_cand}
The observations covered the wavelengths of \lya\ for all but one of
the strong absorption systems listed in Table~\ref{tab:list_obj}. Only
the highest redshift sub-DLA system towards Q2155+1358 was not
covered, i.e. the total number of systems included in this analysis is
21 DLA and sub-DLA systems.

\subsection{Expected sizes}
For this project we are only interested in small wavelength regions
corresponding to \lya\ at the DLA redshifts, and thus the search for
candidate galaxies could be carried out using customised narrow-band
filters.  IFS, on the other hand, has the advantage that the widths of
the narrow-band filters can be adjusted to match those of the emission
lines.  Typically customised narrow-band filters have a larger
transmission \textit{FWHM} than the spectral resolution of IFS data.
Hence, IFS allows detection of emission line objects with a higher
signal-to-noise ratio than that possible with narrow-band filters.  A
disadvantage is the relatively small field of view of current IFUs,
but this is not a serious concern.  One can estimate the expected
sizes of DLA galaxies \citep[see][]{wolfe86}. Using a Schechter
luminosity function and a power-law relation between the disk
luminosity and gas radius given by the Holmberg relation
\(R/R^*=(L/L^*)^{\beta}\), one can calculate the expected impact
parameter. Combining $\beta=0.26$ found for DLA galaxies at $z<1$
\citep{chen05} with the luminosity function in for $z=3$ galaxies
\citep{poli03} one finds $R^*\approx30$~kpc. If DLA galaxies are
similar to or fainter than $L^*$ galaxies this implies that DLA
galaxies at $z>2$ are expected to lie closer than $\sim$4\arcsec\ from
the QSO line of sight. The small field of view of IFUs is therefore
well suited to search for \lya\ emission from DLA galaxies.

The estimated galaxy sizes are highly dependent on the parameters of
the DLA galaxy luminosity and slope $\beta$.  Most probably, high
redshift DLA galaxies are not regular disks like those in the local
universe. Numerical models of DLAs predict that the galaxies are
mostly smaller than 10 kpc, while observations that give limits on the
star-formation rates associated with DLAs suggest that DLAs are
located in neutral gas around Lyman break galaxies \citep{wolfe06}.
As DLA galaxies at $z>2$ are generally found to be fainter than an
$L^*$ galaxy \citep{colbert02}, we choose to consider only objects
with impact parameters smaller than 30 kpc for a more detailed
analysis.

The impact parameters that we measure in the data correspond to the
radially projected distances so the real distances to the absorber can
be larger.  Two candidates are found at impact parameters larger than
30 kpc, and they are likely not associated directly with the absorbers
themselves.

\subsection{Candidate selection}
\label{sect:cand_sel}
Some \lya\ emission lines from DLA galaxies are offset from the QSO-
DLA line by $\sim200$~km~s$^{-1}$ \citep{moller02}, whereas
\lya\ emission from high redshift galaxies can have even larger
offsets from the galaxy systemic redshift \citep{shapley03}. We
therefore chose to focus on regions in the data cubes with velocities
ranging from approximately --1000 to +1000 km s$^{-1}$ from the DLA
lines.

First, the reduced data cubes were stacked in a two-dimensional frame
and inspected visually around the DLA lines for emission line
objects. When the spatial offset from the QSO is larger than the
seeing, or alternatively when the QSO is very faint, 
emission line objects can be identified directly because of the
ordering of the spectra in the stacked spectrum.  Where no objects
could be detected visually further sampling of the data cubes was
necessary to increase the signal-to-noise ratio to detect candidate
emission line objects. Inspections of the data cubes was done using
the Euro3D visualisation tool \citep{sanchez04a}.

From the reduced, sky-subtracted and combined data cubes, narrow-band
images were created with an initial width of 10--15~{\AA} depending on
the spectral resolution of the observations. A set of images was
created offset by --10 to +10~{\AA} from the DLA line to allow for
possible velocity shifts of the \lya\ emission line,
and inspected visually for objects brighter than the background.  If
detected, spectra from these brighter regions were co-added and
inspected for emission lines at the wavelength chosen in the
narrow-band image. This step was necessary to discriminate between
emission lines and individual noisy spectra.  It is known that three
blocks of 16 fibres, i.e. 48 fibres in an area of 1\farcs5 to the west
in the field of view, have lower than average transmission.  The
effect of correcting for the total throughput was that these spectra
had lower signal-to-noise ratios. When narrow-band images were created
from the cubes, the higher variance in these spaxels could result in
extreme values, seemingly inconsistent with the neighboring spaxels.
Only by looking at the spectrum associated with a bright spaxel could
it be determined if an emission line was present, or if the spectrum
was just noisy.  If an emission line was seen, a second pass
narrow-band image was created using the value of the emission line
width to increase the signal of the detection.  A second pass
one-dimensional spectrum was created after inspecting the narrow-band
image for more bright spaxels surrounding the emission line
candidate. This procedure was iterated until the signal in either
narrow-band images or spectra did not increase.  We found that an
interactive visual identification of faint emission lines was more
effective than an automatic routine.

\begin{figure*}[!htpb]
\centering
\resizebox{15cm}{22.3cm}{\includegraphics[bb= 10 35 596 842, clip]{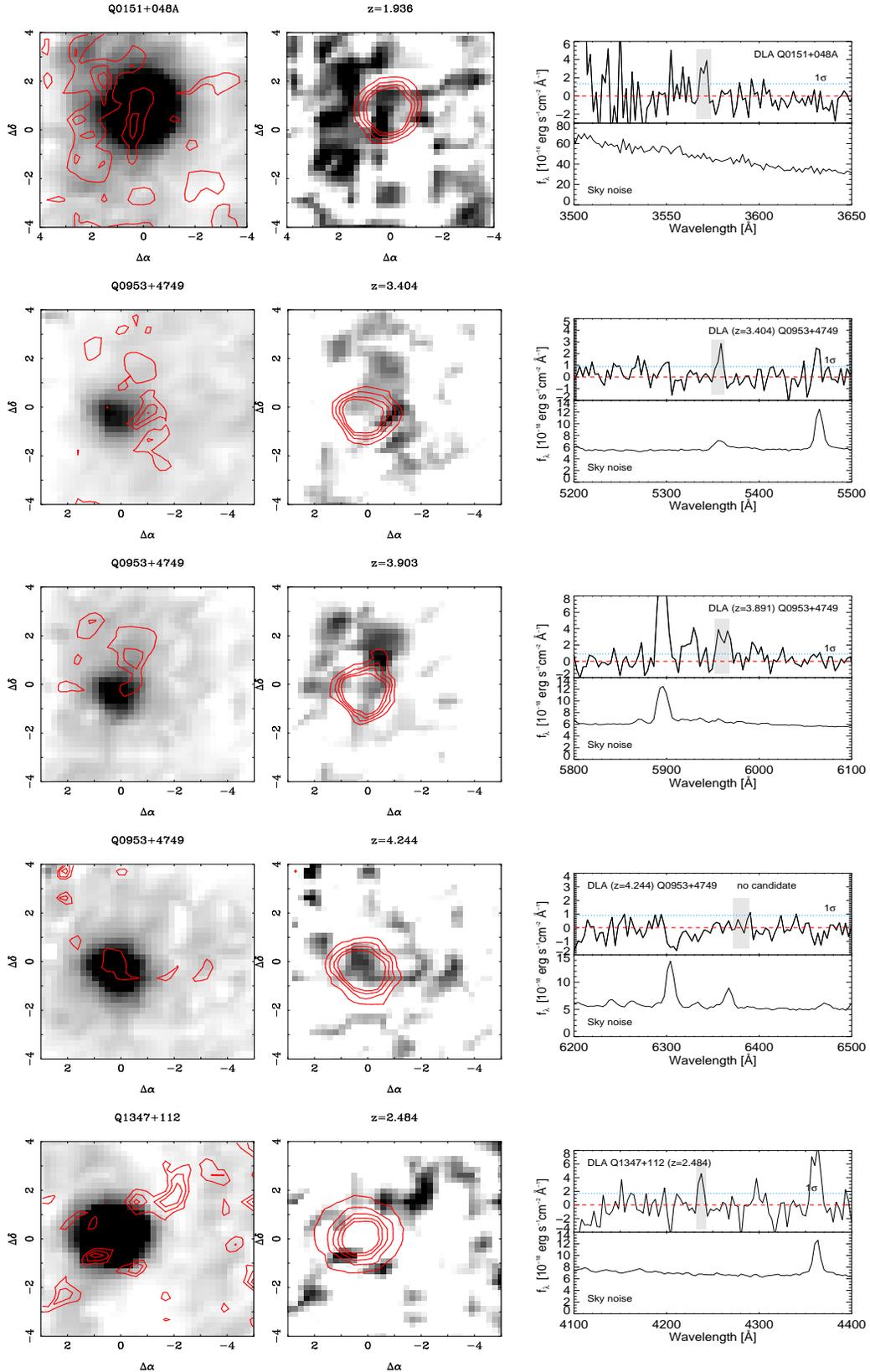}}
\caption{\textit{Left panels}: narrow-band images of the QSOs with
  overlayed contours of narrow-band images centered on the
  \lya\ wavelengths of the DLAs.  The images are 8\arcsec\ square.
  Contour levels correspond to 2, 3, 4$\sigma$ levels above the
  background noise. \textit{Middle panels}: the reverse, where the
  contours are arbitrary apart from the central one that shows the QSO
  seeing \textit{FWHM}. \textit{Right hand panels}: Spectra of
  candidate emission line objects created from co-adding spectra from
  spaxels associated with the emission line candidates. The width of
  the grey bars over the emission lines correspond to the wavelength
  ranges of the narrow band images.  The lines below the spectra show
  the background sky noise spectra determined from background
  spaxels. }
\label{fig:hiz_cand_imspec}
\end{figure*}

To allow a better visual detection of emission line objects, the
narrow-band images were interpolated to pixel scales 0\farcs2
pixel$^{-1}$ as shown in Fig.~\ref{fig:hiz_cand_imspec}. In all panels
the images are 8\arcsec\ by 8\arcsec, with orientation north up and
east left.  The left panels show interpolated images of the QSO at
wavelengths near to the DLA line.  Contours correspond to an image
centered on the visually detected emission feature close to the DLA
redshift.  In the middle panels in Fig.~\ref{fig:hiz_cand_imspec} the
plots are reversed, such that the image shows the emission line object
and the contours correspond to the QSO narrow-band image.  Here, the
innermost contour corresponds to the seeing \textit{FWHM}. To enhance
the visibility of the candidates the QSO emission was subtracted from
the data cubes before creating the images.  This subtraction of the
QSO emission was done using a simple approach
\citep[see][]{christensen06}.  A scale factor was determined for each
spaxel by dividing each spectrum by the extracted one-dimensional QSO
spectrum. Using this scale factor, the QSO emission was subtracted, a
process which retains objects with spectral characteristics different
from the QSO in the data cube.

The spectra of the candidates are shown in the right hand column in
Fig.~\ref{fig:hiz_cand_imspec}. These are created by co-adding spectra
from between 4 and 10 spaxels.  The dotted line corresponds to the
1$\sigma$ noise level determined from a statistical analysis of the
pixel values in the data cube, while the lower sub-panels show the
background noise spectra in the data cubes, obtained from 4-10
background spaxels.

\addtocounter{figure}{-1}
\begin{figure*}[!htpb]
\centering\resizebox{15.5cm}{24.5cm}{\includegraphics[bb= 10 35 596 842, clip]{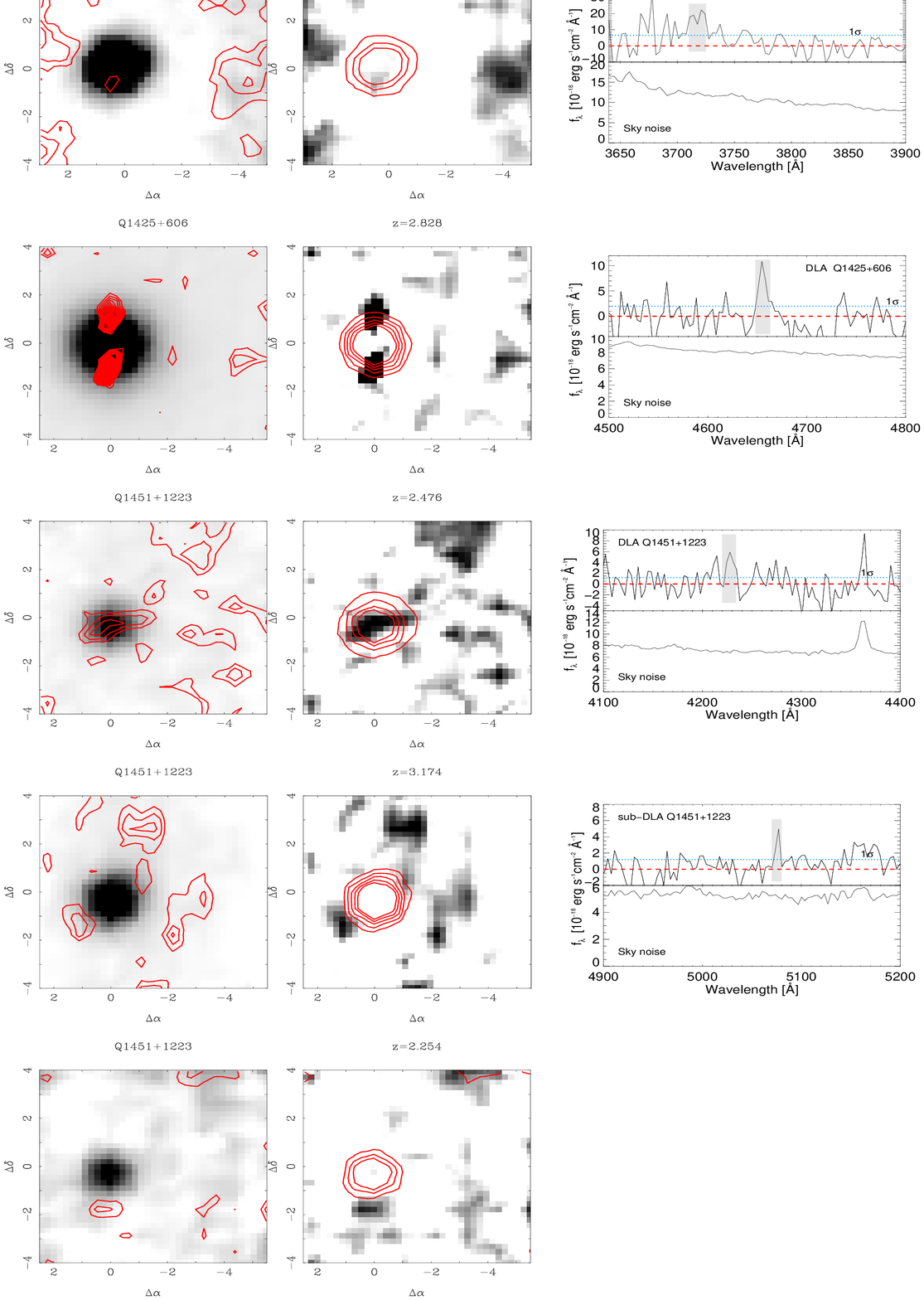}}
\caption{Plots of candidates-- continued. No candidate is found for
  the $z=2.254$ DLA towards Q1451+1223. }
\end{figure*}
\addtocounter{figure}{-1}
\begin{figure*}[!htpb]
\centering\resizebox{15.5cm}{24.5cm}{\includegraphics[bb= 10 35 596 842, clip]{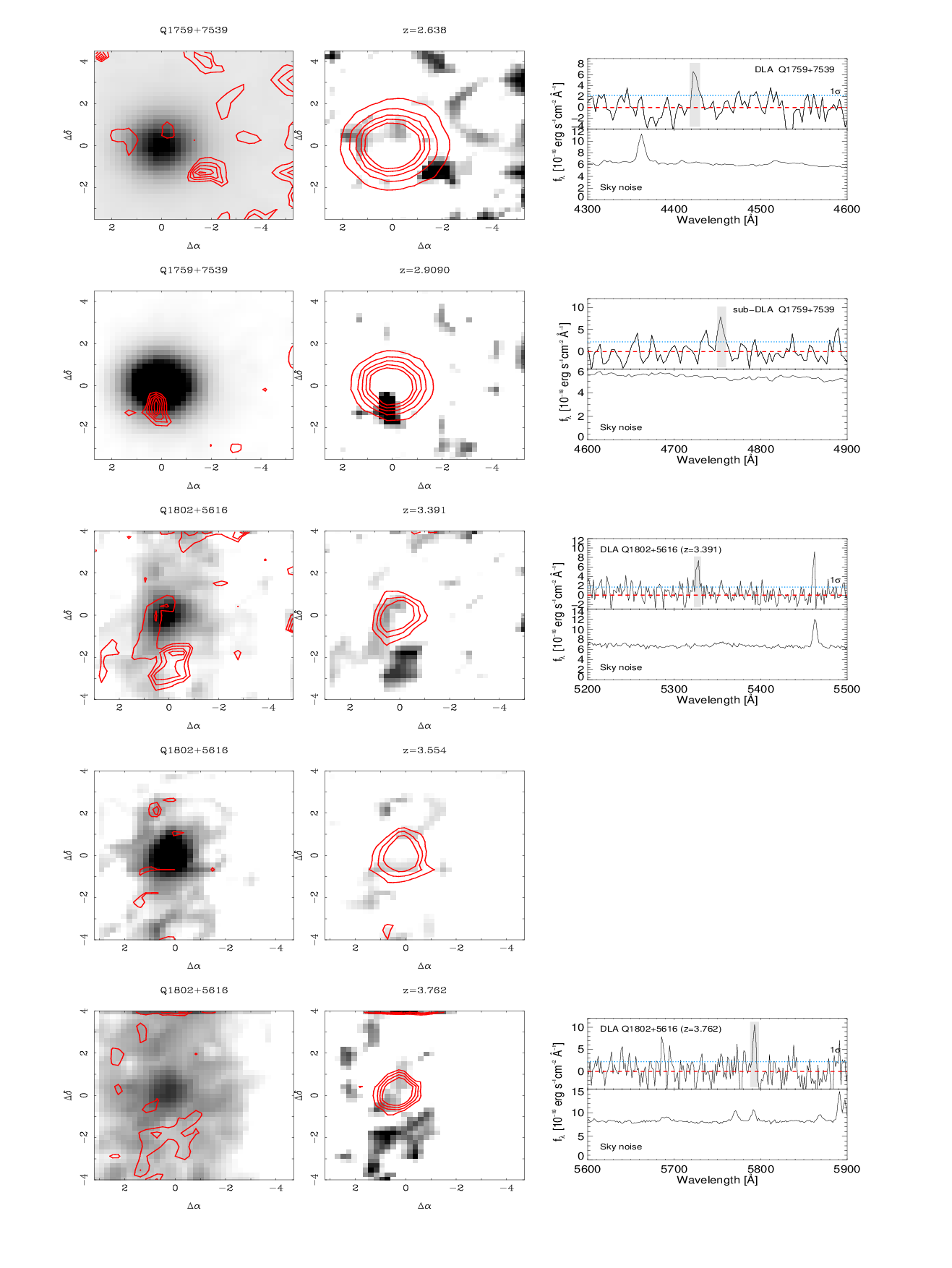}}
\caption{Plots of candidates-- continued. No candidate is found for
  the $z=3.554$ DLA towards Q1802+5616.}
\end{figure*}
\addtocounter{figure}{-1}
\begin{figure*}[!htpb]
\centering\resizebox{15.5cm}{24.5cm}{\includegraphics[bb= 10 35 596 842, clip]{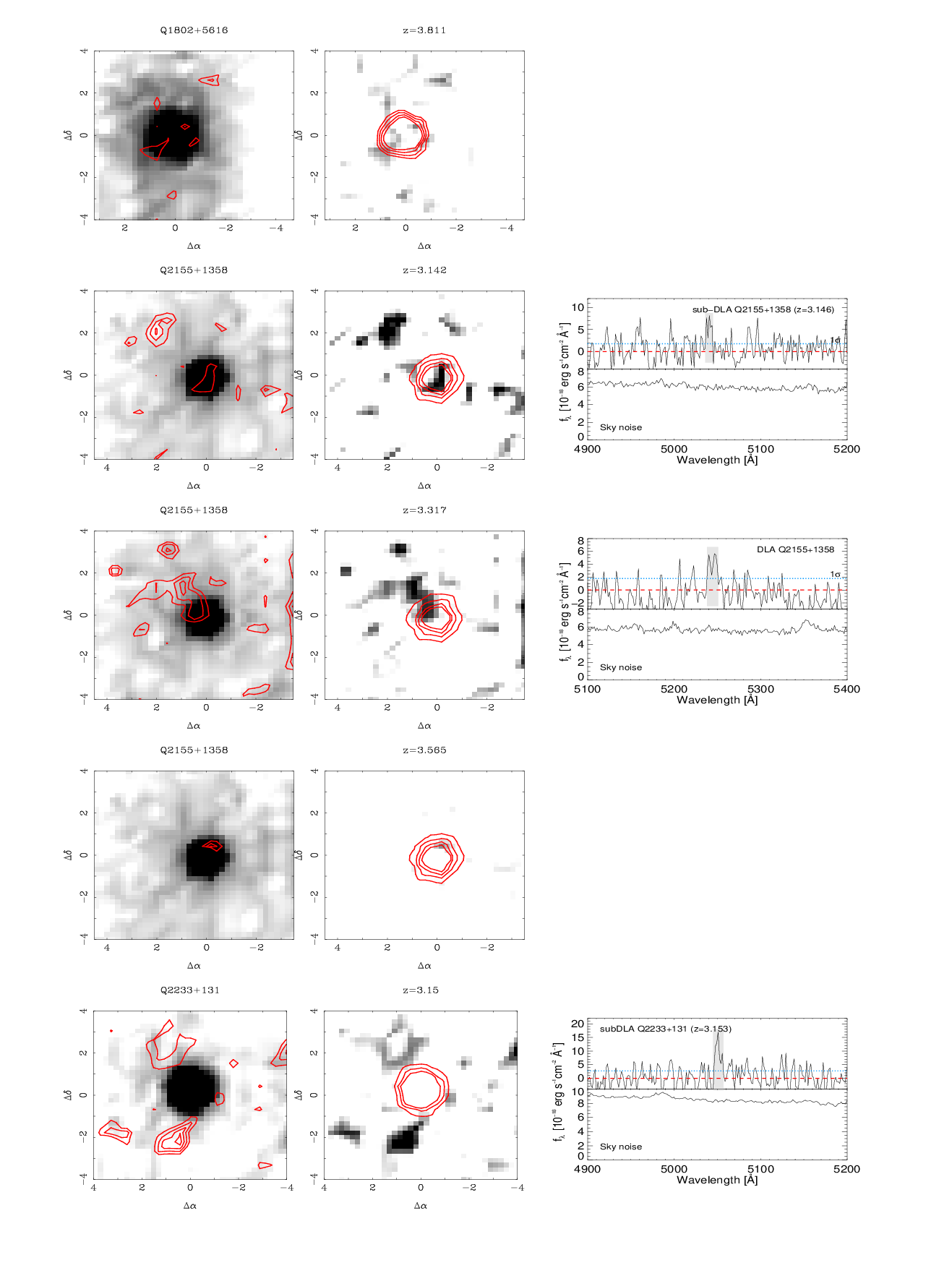}}
\caption{Plots of candidates-- continued. No candidates are found for
  the $z=3.811$ DLA towards Q1802+5616, or the $z=3.565$ sub-DLA
  towards Q2155+1358.}
\end{figure*}
\addtocounter{figure}{-1}
\begin{figure*}[!htpb]
\centering\resizebox{\hsize}{6.cm}{\includegraphics[bb= 10 660 596 842, clip]{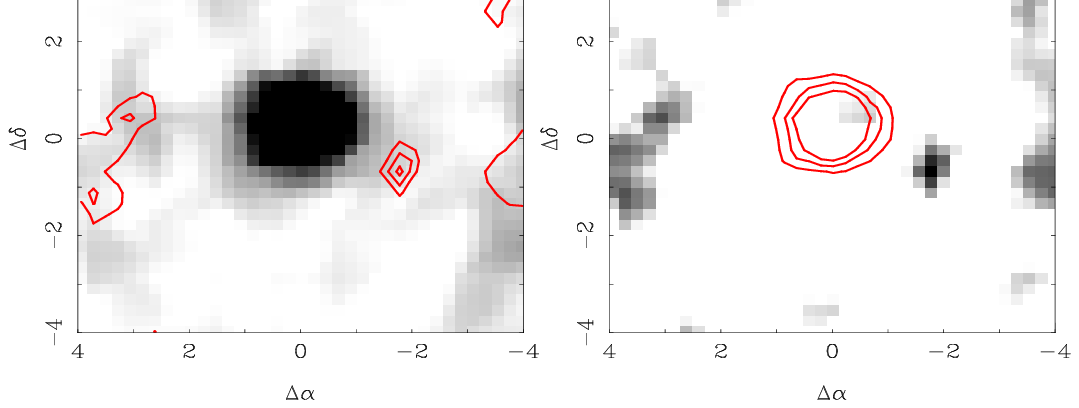}}
\caption{Plots of candidates-- continued. No candidate is found for
  the $z=2.51$ sub-DLA towards Q2233+131.}
\end{figure*}

Properties of the candidate objects corresponding to those with
spectra in Fig.~\ref{fig:hiz_cand_imspec} are listed in
Table~\ref{tab:hiz_cand_prop}.  Offsets in RA, DEC from the QSO and
the corresponding projected distance at the DLA redshift are listed in
columns 2, 3, and 4. Emission lines were fit using {\tt ngaussfit} in
IRAF, redshifts listed in column 5 are derived.  Fluxes in column 6
are derived from the Gaussian fits, and errors in the peak intensity,
line width, and continuum placement are propagated to calculate the
uncertainties. The fluxes have not been corrected for the Galactic
extinction. The flux measurements and the associated errors indicate
that most of the candidates are detected with a signal-to-noise ratio
$<4\sigma$.  Column 7 gives the velocity difference between the
systemic DLA redshift and the candidate \lya\ emission lines.  We
integrate the signal-to-noise estimate over the emission line (Column
8), \(\textrm{S/N}_{int} = f/(\sqrt{N}\times \sigma) \), where $f$ is
the line flux, $N$ is the number of pixels the emission line covers
and $\sigma$ is the noise in adjacent wavelength intervals. Column 9
gives the observed emission line \textit{FWHM} after correcting for
the instrumental resolution.  Finally column 10 gives the significance
classes of the candidate detection, which is explained in
Section~\ref{sect:hiz_signific}.

Columns 3 and 4 in Table~\ref{tab:hiz_cand_sig} list the values of
Galactic reddening towards each QSO \citep{schlegel98}, and the
correction factors to be applied to the candidate fluxes for a Milky
Way extinction curve \citep{fitzpatrick99}.

\begin{table*}[!ht]
\begin{footnotesize}
\centering
  \begin{tabular}{llllllllll}
   \hline \hline
   \noalign{\smallskip}
QSO  &  $\Delta$RA& $\Delta$DEC &  $b$ & $z$ &
   $f_{\lambda}$ &  $\Delta$ v  & S/N$_{int}$ & \textit{FWHM}  & significance\\
     &  (\arcsec) &  (\arcsec) & (kpc) & & & (km s$^{-1}$) & & (km
s$^{-1}$) & class \\
(1)  & (2) & (3) & (4) & (5) & (6) & (7) & (8) & (9) & (10)\\
  \noalign{\smallskip}
   \hline
   \noalign{\smallskip}
Q0151+048A & 2.5:0.4 & --2.5: 1.7 & 3.3:25.4 & 1.9363 & (150$\pm$70) & +225 & 7  & 280$\pm$220 & conf.\\
  \noalign{\smallskip}
Q0953+4749 & --1.2 &  --0.2& 9.0 & 3.4041 & (6.6$\pm$2.9)  & 0  & 15 &   290$\pm$260 & 2 \\
           & --0.5 & 1.8  & 11.1 & 3.9029 & (4.9$\pm$2.1)  & +730& 16&   570$\pm$230 & 3\\
  \noalign{\smallskip}
Q1347+112 & --2.0 & 1.5  & 20.2  & 2.4835 & (3.5$\pm$1.9)  & +1080 & 8 &   190$\pm$370 & 3\\
          & --4.1 & 0.4  & 34.3  & 2.0568 & (4.2$\pm$3.0)   &
   +670  & 7 & 610$\pm$210& 1 \\
  \noalign{\smallskip}
Q1425+606 &  --4.1& 0.4  & 32.3  & 2.8280 & (8.5$\pm$3.1) &
   +80 & 14 & 590$\pm$220 & 3\\
  \noalign{\smallskip}
Q1451+1223 & --3.0 & 3.8 & 39.2  & 2.4764 & (5.8$\pm$2.6)   &
   +640  & 10 & 320$\pm$260 & 3\\
           & --1.0 & 2.8 & 22.5  & 3.1739 & (3.1$\pm$2.0) &
   +210  & 7 & $<$100 & 3\\
  \noalign{\smallskip}
Q1759+7539 & --1.4 & 3.5  & 30.0 & 2.6377 & (5.8$\pm$3.0)   &
   +1050 & 10 & 290$\pm$220 & 2\\
         & 0.1     & --1.2& 9.4  & 2.9090 & (6.0$\pm$2.9)  &
           --80 & 21 & 240$\pm$260 & 1 \\
  \noalign{\smallskip}
Q1802+5616 & --0.2 & --2.0  & 14.9 & 3.3820 & (3.5$\pm$0.9) &
           --610 & 10 & 180$\pm$90 & 4\\
           & 0.2  & 1.8 & 12.9 & 3.7652 & (4.6$\pm$1.9) & +200 & 7 & $<$100 & 2 \\
  \noalign{\smallskip}
Q2155+1358 & 0.7  &  1.2 & 10.2 & 3.3174 & (9.4$\pm$3.0)  &
           +100  & 9 & 780$\pm$210 & 3\\
& 1.7  &  1.9 & 19.4 & 3.1461 & (4.1$\pm$2.4)  &
           +290  & 7 & 260$\pm$220 & 3\\
  \noalign{\smallskip}
Q2233+131 &  0.6  &  2.3 & 18.0 & 3.1543 & (9.6$\pm$2.5)    &
           +90  & 9 & 230$\pm$110 & conf.\\
  \noalign{\smallskip}
   \hline
  \end{tabular}
  \caption[]{Properties of candidate \lya\ emission lines. Column 2,
    3, and 4 list the offsets of the candidate in RA and DEC and in
    projected kpc at the \lya\ emission redshifts (in col. 5),
    respectively. Column 6 lists the integrated \lya\ flux in units of
    $10^{-17}$~\ecs, and column 7 the velocity offset from the DLA
    redshift. Column 8 lists the integrated signal-to-noise ratio of
    the \lya\ emission line, and column 9 gives the line width of the
    emission lines. Fluxes have not been corrected for Galactic
    extinction. Column 10 lists the significance class of the
    detections as described in Sect.~\ref{sect:hiz_notes}. `Conf.'
    implies candidates that were confirmed previously
    \citep{moller98,djorgovski96}.}
  \label{tab:hiz_cand_prop}
\end{footnotesize}
\end{table*}

\section{Results}
\label{sect:hiz_results}
This section describes the classification of the emission line
candidates. We estimate the contamination from spurious detections and
from interlopers. Notes on each observed object are presented as well.

\subsection{Candidate significance class}
\label{sect:hiz_signific}
To estimate how reliable the candidate detection was, various tests
were applied to the data cubes. The candidates were assigned a
significance class: 1, 2, 3, and 4 according to how many of the
following tests were passed.
\begin{enumerate}
\item Instead of co-adding all data-cubes, two independent subsets of
  exposures were created, and the emission line candidate was visible
  in both sub-set combinations. In Sect.~\ref{sect:hiz_notes} these
  will be referred to as subcombinations.
    
\item The emission line candidate was visible in the `simple
  extractions', i.e. where a Gaussian profile was not assumed. 
    
\item Emission line candidates were visible in the narrow-band images
  when the QSO spectrum was subtracted from the data cube.
    
\item Emission line objects that were directly visible in individual
  or combined data cubes, or in the stacked spectra.
\end{enumerate}
In all cases, for a candidate to be considered further it was required
to be detected above 3$\sigma$ in both narrow-band images and the
associated spectra.  The significance classes of the candidates are
listed in column 10 in Table~\ref{tab:hiz_cand_prop}, and comments for
each object are described in Section~\ref{sect:hiz_notes}.  Since the
candidates were found from visual inspections of the data cubes, this
classification was done to describe the candidates in a more
qualitative manner. As the classes involve various tests on the data
sets, this classification goes beyond the simple statistical
significance in terms of to what $\sigma$ level the object is
detected.

\begin{table}
\begin{footnotesize}
\centering 
\begin{tabular}{lll}
 \hline \hline
   \noalign{\smallskip}
QSO  & \ebv & $f_{\mathrm{frac}}$\\
  \noalign{\smallskip}
   \hline
   \noalign{\smallskip}
Q0151+048A              &0.044&1.216\\
Q0953+4749 ($z=$3.4041) &0.011&1.032\\
Q0953+4749 ($z=$3.9028) &     &1.028\\
Q1347+112  ($z=$2.4835) &0.035&1.145\\
Q1347+112  ($z=$2.0568) &     &1.163\\
Q1425+606  ($z=$2.827)  &0.012&1.043\\
Q1451+1223 ($z=$2.4764) &0.031&1.128\\
Q1451+1223 ($z=$3.1739) &     &1.102\\
Q1759+7539 ($z=$2.6377) &0.053&1.220\\
Q1759+7539 ($z=$2.91)   &     &1.199\\
Q1802+5616 ($z=$3.3820) &0.052&1.164\\
Q1802+5616 ($z=$3.7652) &     &1.145\\
Q2155+1358 ($z=$3.3174) &0.067&1.222\\
Q2155+1358 ($z=$3.1461) &0.067&1.237\\
Q2233+131               &0.068&1.240\\  
   \noalign{\smallskip}
\hline
\end{tabular}
\caption[]{ Column 2 and 3 give values of the Galactic reddening and
   the corresponding correction factor to be applied to the emission
   line candidates.}
  \label{tab:hiz_cand_sig}
\end{footnotesize}
\end{table}

\subsection{Non-detections}
In the data cubes where no candidates were found, we estimated the
upper limits for the emission line fluxes.  Spectra from spaxels
within one seeing element (e.g. 4 spaxels corresponding to a seeing of
1\arcsec) were co-added to create a one-dimensional spectrum.
Artificial emission lines with varying line fluxes were added to this
spectrum at the DLA wavelength, and Gaussian profile fits to these
lines were used to estimate the detection level.  The results are
listed in Table~\ref{tab:hiz_non_det}.  The varying limits are due to
the wavelength dependent noise in the data cubes and in particular the
presence of residuals from nearby sky emission lines.

\begin{table}
\begin{footnotesize}
\centering 
\begin{tabular}{lll}
 \hline \hline
   \noalign{\smallskip}
QSO & $z_{\mathrm{abs}}$ & $f_{\mathrm{lim}} (3\sigma)$\\
  \noalign{\smallskip}
   \hline
   \noalign{\smallskip}
Q0953+4749 & 4.244 & 2.5\\
Q1451+1223 & 2.256 & 4.0\\
Q1802+5616 & 3.554 & 4.0\\
Q1802+5616 & 3.811 & 2.2\\
Q2155+1358 & 3.565 & 3.5\\
Q2233+131  & 2.551 & 4.8\\
   \noalign{\smallskip}
\hline
\end{tabular}
\caption[]{DLA and sub-DLA systems where no candidate emission lines
  are found and 3$\sigma$ upper limits for the line fluxes.  Fluxes
  are in units of $10^{-17}$ \ecs.}
  \label{tab:hiz_non_det}
\end{footnotesize}
\end{table}

\subsection{Experiments with artificial objects}
\label{sect:hiz_art}
To investigate how the efficiency of the visual inspection depended on
object properties, several experiments with artificial data cubes were
made. Similar to artificial experiments for one- and two-dimensional
data sets, artificial emission line objects were added to the data
cubes.  These objects were described by the location in RA and DEC,
central wavelength, peak emission intensity, and the widths in RA, DEC
and wavelength. For simplicity we assumed that an emission line object
seen as a point source in the data cube could be represented by a
Gaussian profile in each direction, i.e. described by a Gaussian
ellipsoid in the data cube. 

We first tested completely simulated data cubes with statistical noise
levels corresponding to the typical noise level in the combined data
cubes.  An emission line object with a flux of
$5~\times~10^{-17}$~\ecs, a width of 800 km s$^{-1}$, and spatial
\textit{FWHM} of 1\arcsec\ was placed at a previously known
wavelength.  In the stacked spectra no objects could be seen
immediately. The emission line was only identified after inspecting
the data cube in the visualisation tool, and it was extracted and
analysed in the same way as the real data.  Similar tests were made by
adding an emission line to a real data cube, where the background
noise included the systematic noise as well as the pure Poissonian
noise.  These tests produced similar results for the faint emission
lines with \lya\ flux $f\sim5\times10^{-17}$ \ecs, i.e. 1) the
emission line flux could be recovered within uncertainties, 2) even at
very small impact parameters the object could be found 3) the
reconstructed PSF of the emission line object was irregular as in any
of the images in Fig.~\ref{fig:hiz_cand_imspec}.

We also tested an automatic routine where the re-detection of the
artificial objects was done with no visual intervention.  A set of
narrow-band images were created in wavelength ranges around the
artificial line. For the detection of an emission line the location
was constrained to be within $\pm$10 {\AA} of the input central
wavelengths.  These images were smoothed and a two-dimensional
Gaussian profile was fit to the images. When an object was detected
above a certain threshold, spaxels around the centre within the seeing
\textit{FWHM} were co-added. A series of tests showed that the
recovered flux was consistent within 1$\sigma$ errors for fluxes down
to $f=5\times10^{-17}$ \ecs.  In a typical data cube this was also the
detection limit where 50\% of the objects were re-identified, while
the fraction of re-identified emission lines at this flux level from a
visual inspection was larger.

Tests on the frequency of false detections in data cubes where no
objects were present showed that simultaneous detections of objects in
narrow-band images and associated spectra with S/N~$>$~3 occurred at a
rate of less than 5\% in a series of experiments. Therefore false
detections cannot explain the large number of candidate objects.

\subsection{Field \lya\ emitters}
We estimate here whether the detected candidates are likely to be
field \lya\ emitters having no association with the DLAs.
Observations of high redshift objects have partly focused on detecting
\lya\ emission from galaxies to determine the global comoving
star-formation rates \citep[e.g.][]{hu98,hu04}.  The density of
\lya\ emitters at $z\sim3$ is estimated to be 15000 deg$^{-2}$ $\Delta
z^{-1}$ with line fluxes brighter than a mean of $f=1.5\times10^{-17}$
\ecs\ \citep{hu98,kudritzki00}. From the luminosity function at
$z\approx3$ \citep{vanbreukelen05}, the expected number of field
\lya\ emitters at a flux limit of $5\times10^{-17}$ \ecs\ is
1.7$\times$10$^{-4}$ arcsec$^{-2}$ $\Delta z^{-1}$.  In our survey,
the 9 data cubes sample a total redshift interval of $\Delta z=21.55$
around $z\approx3$.  Statistically, it is expected that there are 0.2
field \lya\ emitters in the whole sample presented here.  Because
these very faint lines are difficult to locate when the approximate
wavelength is not known in advance, we did not look for field emission
objects.  The negligible number of expected field emitters furthermore
shows that the emission candidates, if proved to be real, are unlikely
to be interloping field \lya\ emitters. They are much more likely to
be associated with the DLA galaxies.

\subsection{Notes on individual objects}
\label{sect:hiz_notes}
This section explains the significance of the candidates for each
individual QSO.

{\it Q0151+045A. --} This is a $z_{\mathrm{em}}\approx
z_{\mathrm{abs}}$ system at $z\approx1.93$.  After flux calibration,
the QSO spectrum is 2 magnitudes brighter than that presented in
\citet{moller98}. The low instrument sensitivity at 3560~{\AA}
combined with a variable extinction coefficient at Calar Alto makes
the calibration uncertain.

Extended \lya\ emission was observed in a region of
$3$\arcsec$\times6$\arcsec\ around the QSO mostly to the east of the
QSO \citep{fynbo99}. Long slit spectroscopy along the long axis
revealed velocity structures of 400 km s$^{-1}$ that could be
interpreted as a rotation curve \citep{moller99}.  In the IFS data
extended emission is detected to some degree in
Fig.~\ref{fig:hiz_cand_imspec}, but not with the same detail as in the
higher spatial resolution and larger field of view data in
\citet{fynbo99}. This is the only case in the sample where extended
emission is found, but the signal is not strong enough to determine
the velocity structure over the extended region. The spectrum shown in
Fig.~\ref{fig:hiz_cand_imspec} is the total spectrum co-added from the
whole nebula.

{\it Q0953+4749}. -- This $z_{\mathrm{em}}=4.457$ QSO has three DLAs
at $z_{\mathrm{abs}}$~=~ 3.407, 3.891, and 4.244 \citep{bunker03}.  A
candidate associated with the lowest redshift DLA is visible in the
narrow-band image in Fig.~\ref{fig:hiz_cand_imspec}. Independent
subcombinations, the simple extraction, and the corresponding spectra
show a faint emission line.  This emission line coincides with a sky
emission line 1.6~{\AA} away and could be due to sky subtraction
errors, so we only assign this candidate a significance class of 2. A
\lya\ emission line from the DLA galaxy has been reported (A. Bunker,
private comm.) but its line flux is below our detection limit. For the
second DLA system at $z=3.891$ the object is present in
subcombinations, the simple-extracted images, and in the extracted
spectra. This candidate is assigned significance class 3. No candidate
is found for the highest redshift DLA to the detection limit reported
in Table~\ref{tab:hiz_non_det}.

The locations of the candidates are compared to WFPC2 images obtained
from the HST archive, but no continuum counterpart could be
identified.

{\it Q1347+112. --} This $z_{\mathrm{em}}=2.679$ QSO has a DLA at
$z_{\mathrm{abs}}=2.471$ and another possible one at
$z_{\mathrm{abs}}=2.05$, which needs confirmation from spectroscopy at
higher spectral resolution.  An emission candidate for the $z=2.471$
DLA is visible in the subcombinations and the extracted
one-dimensional spectra. In the simple extraction, the spectrum has a
low signal-to-noise ratio and the emission feature in the spectrum is
faint.  We assign a significance class of 3 to this candidate. For the
$z=2.0568$ DLA system we detect a candidate emission line object, but
note an increase in the background noise shortwards of 3750 {\AA}. The
object is not seen in one of the subcombinations, nor the extracted
spectra, and therefore the candidate is assigned a significance class
of 1.

A snapshot WFPC F555W image \citep{bahcall92} obtained from the HST
archive has a 5$\sigma$ limiting magnitude of 24.4 mag arcsec$^{-2}$,
but no continuum counterpart can be seen at the location of the
candidate.

{\it Q1425+606. --} This $z_{\mathrm{em}}=3.163$ QSO has a DLA at
$z_{\mathrm{abs}}=2.827$.  Because this QSO is very bright, strong
residuals within 1\arcsec\ from the QSO centre are present in the
narrow-band image where the QSO emission is subtracted. A faint object
offset by $\sim$4\arcsec\ to the west is visible in the narrow-band
image in Fig.~\ref{fig:hiz_cand_imspec}. The candidate is present in
subcombinations and in the constructed spectra and is assigned a
significance class of 3.  In PMAS data cubes, spaxels in the west
region are more noisy than the average due to an overall lower
transmission. Note that the tests suggest a good candidate, but the
impact parameter ($>30$ kpc) is large.

{\it Q1451+1223. --} 
This $z_{\mathrm{em}}=3.246$ QSO has two DLAs at
$z_{\mathrm{abs}}=2.469$ and $z_{\mathrm{abs}}=2.254$ and a sub-DLA at
$z_{\mathrm{abs}}=3.171$.  For the DLA system at $z=2.469$ an object
appears after the QSO subtraction close to the centre.  It is caused
by residuals, since the spectrum has no emission lines at the expected
wavelength. For the same DLA, a region $\sim$4\arcsec\ to the north
west appears in both narrow-band imaging, subcombinations, simple
extractions, and the constructed spectra.  We therefore assign a
significance class of 3 to it, but note that is has a large impact
parameter (39 kpc). For the $z=3.171$ sub-DLA an object is detected to
the north. Narrow-band images from subcombinations, and simple
extractions show the emission line candidate, but the corresponding
spectra have emission lines with very low signals. The candidate is
assigned a significance class of 3. No candidate is found for the
$z=2.254$ DLA.

A deep optical broad-band image of the field surrounding this QSO was
obtained by \citet{steidel95}, who found no obvious candidates to the
absorbers. \citet{warren01} found one candidate offset by 3\farcs9 to
the south-west of the QSO in a NICMOS image, but this object is
outside the field of view of the IFS data. An HST/STIS archive image
shows that the emission line candidates lie in regions where no
continuum emitting counterpart is found.

{\it Q1759+7539. --} This $z_{\mathrm{em}}=3.05$ QSO has a DLA at
$z_{\mathrm{abs}}=2.625$ and a sub-DLA at $z_{\mathrm{abs}}=2.91$.
The candidate detected for the DLA system in
Fig.~\ref{fig:hiz_cand_imspec} lies near the northern edge of the
field of view and can be affected by flat field errors. Although it is
bright, the candidate is not visible in both subcombinations, and is
therefore assigned a significance class of 2.  A bright area 1\farcs8
south west of the QSO appears after the QSO emission is subtracted but
it is likely due to residuals. It has no emission lines at the
expected wavelength and is not considered further. The higher redshift
sub-DLA system has an emission line candidate which is visible only
after the QSO PSF has been subtracted from the final cube.  However,
the candidate is only visible in one out of two subcombinations and we
assign this candidate a low significance class of 1.

A NICMOS snapshot image showed no bright galaxies near the QSO to a
limit corresponding to an $L^*$ galaxy \citep{colbert02}.

{\it Q1802+5616. --} This $z_{\mathrm{em}}=4.158$ QSO has four DLAs at
$z_{\mathrm{abs}}$~=~3.391, 3.554, 3.762, and 3.811.  The candidate
for the lowest redshift DLA system is directly visible in the reduced
and combined data cube when looking at the stacked spectra. The
candidate can also be identified in individual subcombinations and in
the simple extracted spectrum. Therefore this candidate is assigned
the class 4. In a narrow-band image at the wavelength of \lya\ at
$z=3.7652$ there is an emission region to the south (see
Fig.~\ref{fig:hiz_cand_imspec}) and the corresponding spectrum shows
an emission feature.  However, this line is coincident with a faint
sky emission line, so this candidate is assigned the class 2.  No
candidates are found for the other two DLA systems.

{\it Q2155+1358. --} This $z_{\mathrm{em}}=4.256$ QSO has a DLA at
$z_{\mathrm{abs}}=3.316$ and three sub-DLAs at
$z_{\mathrm{abs}}=$~3.142, 3.565, and 4.212. The observations only
cover \lya\ for the three lower redshift systems.  IFS covering the
highest redshift system has revealed a possible faint candidate
emission line object \citep{francis06}.  The candidate \lya\ emission
line associated with the DLA system is visible in independent
subcombinations and in the simple extraction and is therefore assigned
a high value of 3.  Because of the partial spatial overlap with the
QSO, the emission from the QSO is subtracted to give a cleaned
emission line object and the associated spectrum shown in
Fig.~\ref{fig:hiz_cand_imspec}.  A candidate is found to the south for
the $z=3.142$ sub-DLA system. This object is visible in the simple
extraction, and subcombinations, but only one associated spectrum
shows a clearly detected emission line.  We assign a significance
class of 3 to this candidate. No candidate is found for the $z=3.565$
sub-DLA.

{\it Q2233+131. --} This $z_{\mathrm{em}}=3.295$ QSO has two sub-DLAs
at $z_{\mathrm{abs}}=3.153$ and $z_{\mathrm{abs}}=$~2.551.  The galaxy
responsible for the $z=3.153$ DLA was found by \citet{steidel95}, and
follow-up spectroscopy confirmed this by the detection of
\lya\ emission \citep{djorgovski96}. Previous IFS of this object
suggested that the \lya\ emission was extended
\citep{christensen04a}. This is not confirmed by the higher spectral
resolution data included in this paper, although there appears to be
some faint emission to the east of the object in
Fig.~\ref{fig:hiz_cand_imspec}. The new data and improved data
reduction which optimises the signal-to-noise ratio, confirm the line
flux \lya\ line flux reported in \citet{djorgovski96}. No candidate
was found for the $z=2.551$ sub-DLA system, consistent with the upper
limit from a deeper Fabry-Perot imaging analysis \citep{kulkarni06}.

\section{Properties of candidate DLA counterparts}
\label{sect:hiz_proper}
We proceed with a more detailed analysis of the properties of the
detected candidate \lya\ emission lines. Only those candidates
assigned values 3 and 4 are included. Of the eight good candidates, we
reject two due to their large impact parameters ($>30$~kpc). However,
since they fulfill the criteria for good candidates, they could
instead belong to a brighter component in a group.  The average
redshift of all the DLAs in the whole sample is
$\bar{z}_{\mathrm{sample}}=3.13$ while that of the six remaining
candidates is $\bar{z}_{\mathrm{cand}}= 3.23$, hence we find no
preference for detection of either lower or higher redshift
candidates. We emphasise that the candidates emission lines have
fluxes that are detected at the 3$\sigma$ level, but with this in mind
we compare their properties with those of confirmed \lya\ emission
lines from DLA galaxies.

\subsection{Line fluxes}
\begin{figure}
\centering
\resizebox{\hsize}{!}{\includegraphics[bb= 0 0 566 425,clip]{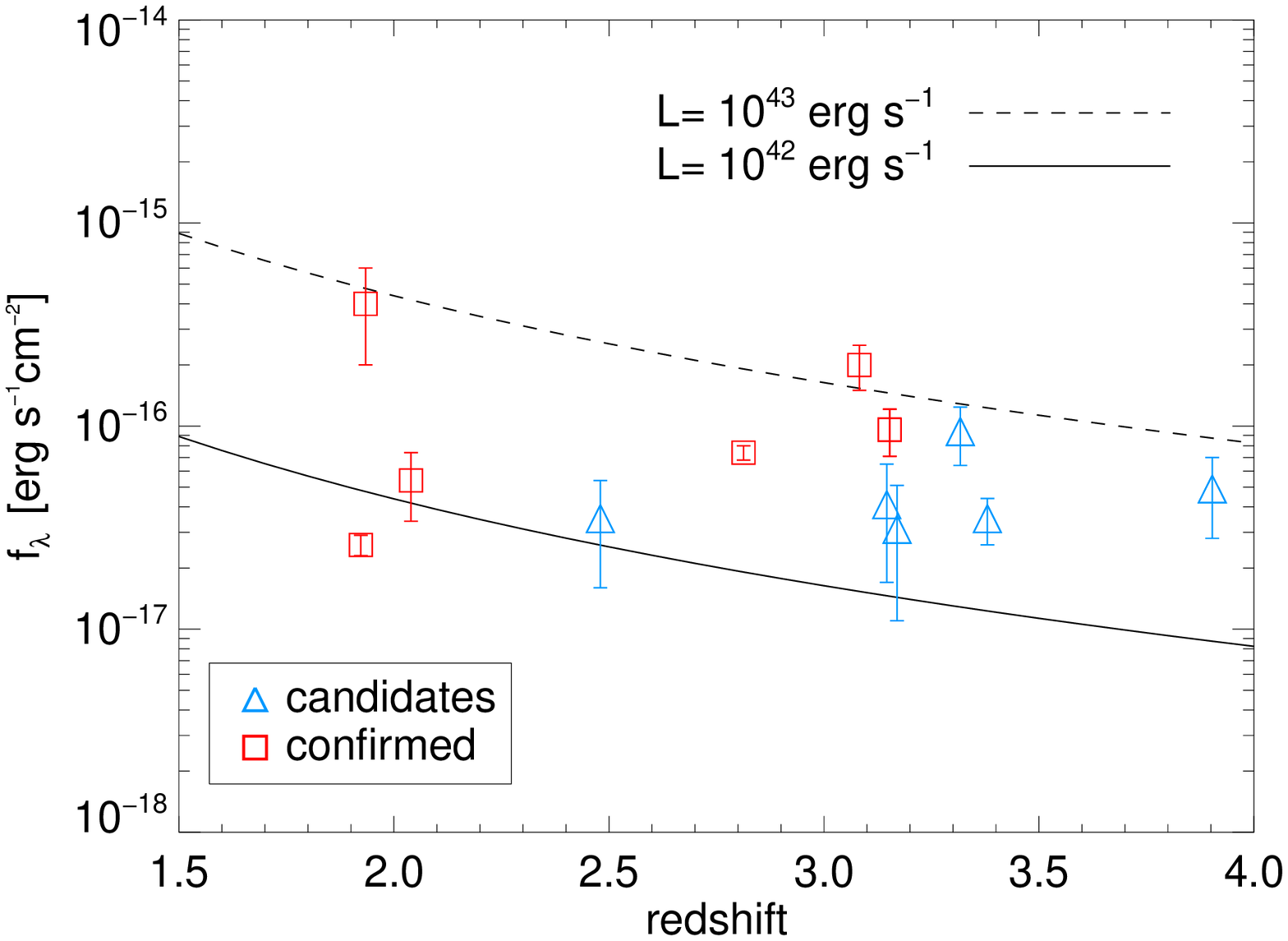}}
\caption{Line fluxes of \lya\ emission objects as function of
  redshift, where square symbols represent already confirmed objects
  and triangles the candidates. The solid and dashed lines correspond
  to \lya\ luminosities of $10^{42}$ and $10^{43}$ erg s$^{-1}$,
  respectively.}
\label{fig:hiz_lum}
\end{figure}

Fig.~\ref{fig:hiz_lum} shows the inferred line fluxes of the
candidates as a function of redshift. The triangles denote our
candidates and square symbols indicate already confirmed objects from
the literature
\citep{moller93,djorgovski96,moller98,leibundgut99,moller02,moller04}. This
figure shows that the line fluxes for the candidates are similar to
those for the previously confirmed ones, which have deeper
observations and detection levels of 5--10$\sigma$.

Fabry-Perot imaging studies of QSOs with DLAs have managed to reach
similar or lower flux limits than our IFS survey
\citep{lowenthal95,kulkarni06}. With their detection limit some
objects should have been detected if the \lya\ fluxes of DLA galaxies
are around the level we find for the candidates and the confirmed
objects. IFS is useful to look for emission lines as it allows us to
adjust a posteriori the central wavelength, whereas in Fabry-Perot
images, the emission line could fall at the wings of the filter where
the transmission is lower.  Another advantage of IFS observations is
the knowledge of the spatial QSO PSF as a function of wavelength which
allows a modeling and subtraction of the QSO emission
\citep{wisotzki03,sanchez04b}. This allows detection of emission lines
even when they are superimposed on the QSO. Nevertheless we do not
detect emission line candidates closer than about 1\arcsec\ from the
QSO possibly due to subtraction residuals. The fact that the confirmed
objects are found at smaller impact parameters compared to the
candidates (Sect.~\ref{sect:hiz_imp}) could indicate a bias.

\subsection{Velocity differences}

An anticorrelation is expected between the \lya\ luminosity and the
velocity difference between the \lya\ emission line and optical
emission lines \citep{weatherley05}. The resonant nature of
\lya\ causes a shift of the emission line towards slightly longer
wavelengths where the photons can escape absorption.  When a larger
fraction of the blue part of the line profile is absorbed, the
remaining emission line of lower luminosity will be more shifted in
velocity compared to brighter ones. This explanation is supported by
the study of \lya\ emission lines from Lyman break galaxies (LBGs)
\citep{shapley03}.

\begin{figure}
  \centering \resizebox{\hsize}{!}{\includegraphics[bb= 0 0 566 425,clip]{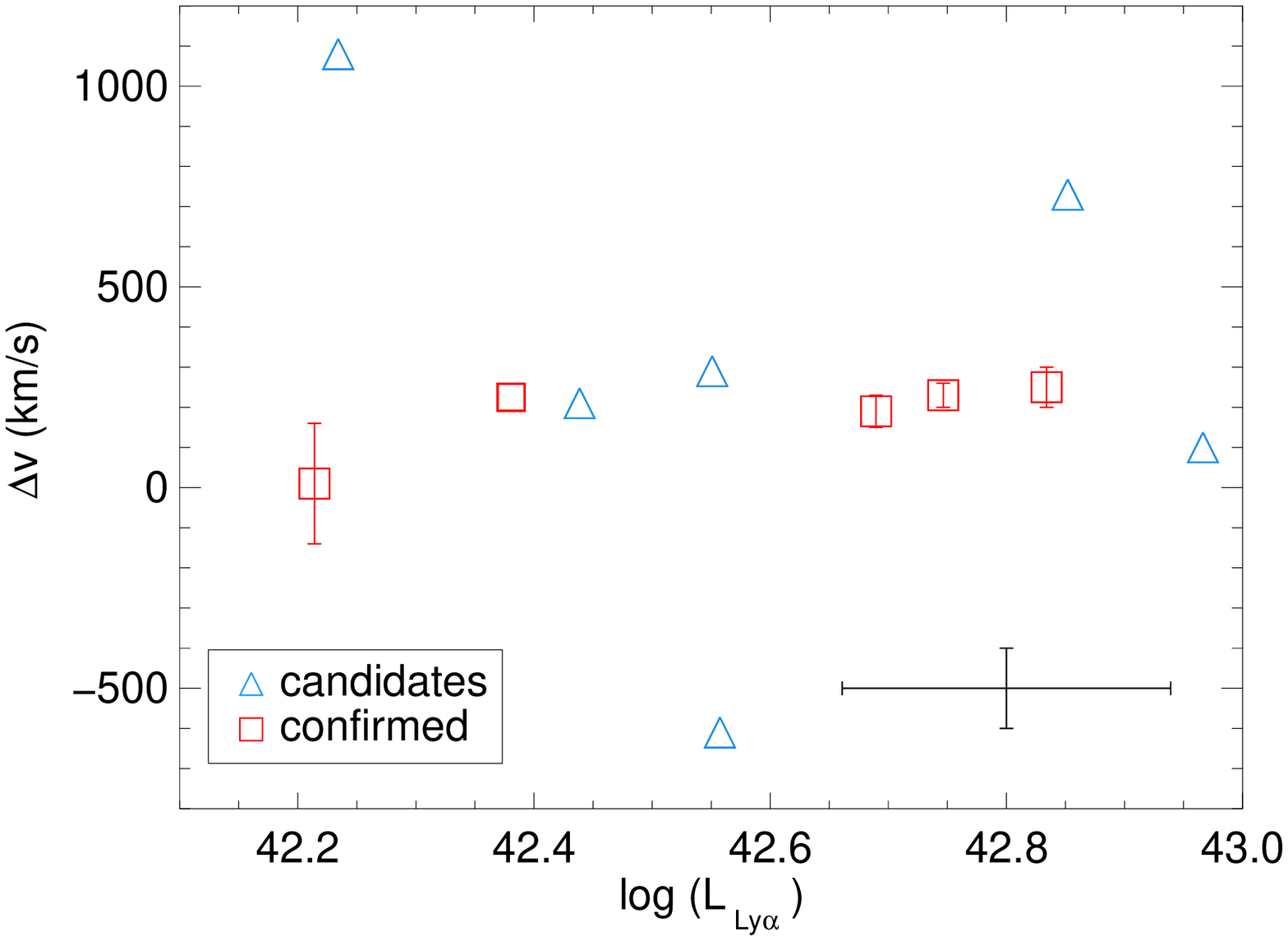}}
\caption{Velocity differences between the \lya\ emission lines and DLA
  redshifts as a function of the \lya\ luminosity. An average error
  bar for the \lya\ emission candidates is shown in the lower right
  corner. }
\label{fig:hiz_lya_vel}
\end{figure}

Fig.~\ref{fig:hiz_lya_vel} shows the velocity differences between
\lya\ emission lines and the DLA redshifts for the candidates as a
function of the \lya\ luminosity. There is no evidence for a
correlation for the candidates. We note that the only candidate that
shows a negative velocity offset is the best candidate in the sample;
the $z=3.391$ DLA towards Q1802+5616. For the candidates we find an
average velocity difference of 300$\pm$580~km~s$^{-1}$, which is
similar to the velocity differences measured for LBGs;
\citet{pettini01} find 560$\pm$410 km~s$^{-1}$ while a larger sample
has $\Delta$v~=~650~km~s$^{-1}$ between the \lya\ emission line and
low-ionisation absorption lines \citep{shapley03}.  In the case that
DLAs are associated with bright galaxies we would expect to see large
velocity offsets too. Furthermore, as the line of sight towards the
emission line object and the QSOs differ by 10--30 kpc, a larger
velocity offset can be expected due to differences in kinematics
within the host and its environment. Instead, if the DLA galaxy
resides in a group, the velocity offset will reflect the velocity
dispersion in the group instead of being related to the host. In
support of this idea, it has been shown that bright Lyman break
galaxies at $z>2$ are surrounded by gas extending to large distances
\citep{adelberger05}. Correlation studies have revealed that DLAs
cluster on almost the same scale as LBGs \citep{cooke06}, indicating
that a similar amount of gas is present in their environments.

  Like flux-limited surveys, this IFS study selects the brightest
  emission component, and it is possible that the real absorbing
  galaxy is a fainter component in a group.

\begin{figure}
  \centering \resizebox{\hsize}{!}{\includegraphics[bb= 0 0 586 425,clip]{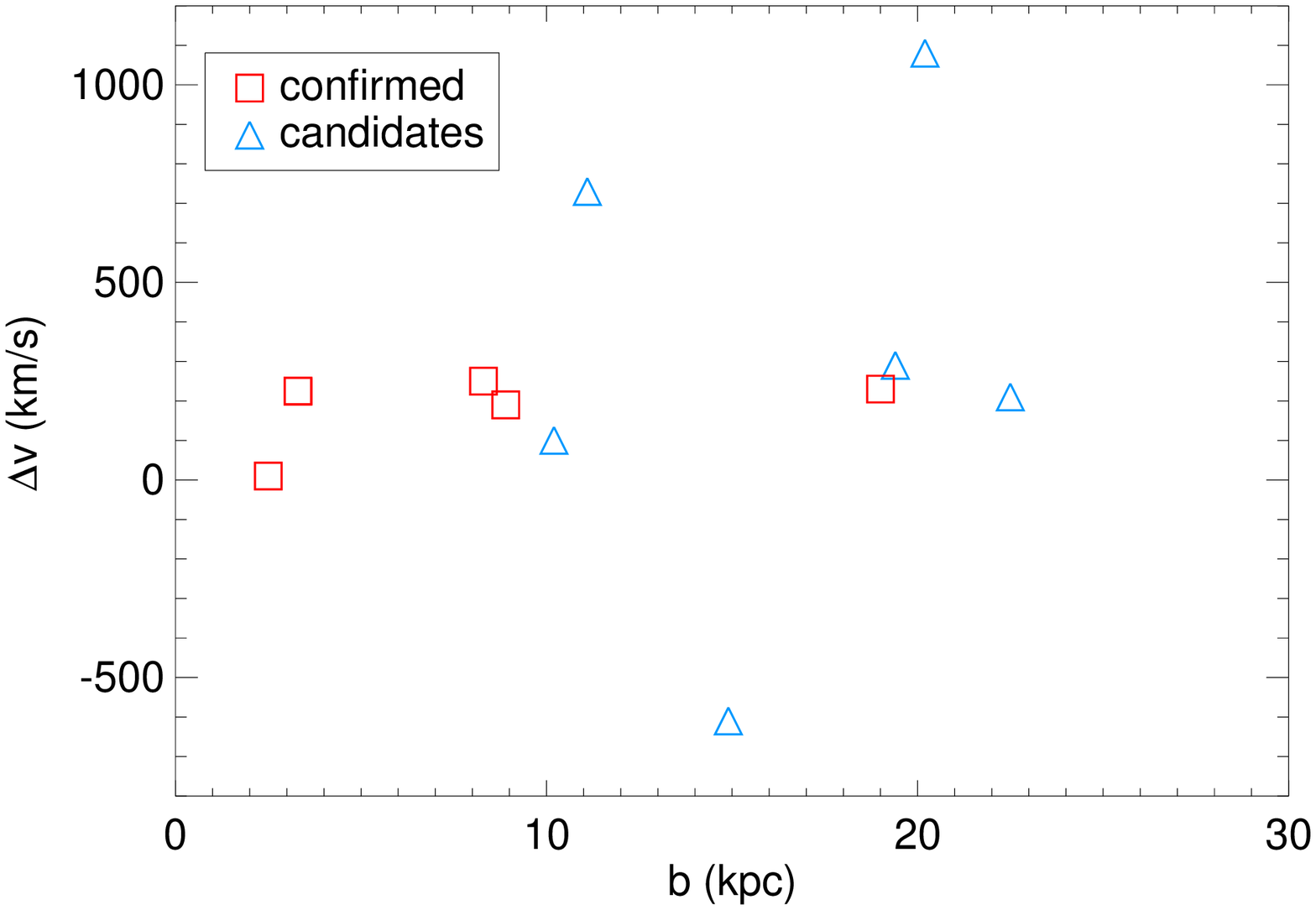}}
\caption{Velocity differences between the \lya\ emission lines and DLA
  redshifts as a function of the impact parameter. }
\label{fig:hiz_vel_impact}
\end{figure}

In the case that DLA galaxies are related to rotating large disks, it
can be expected that the velocity difference increases with impact
parameter but Fig.~\ref{fig:hiz_vel_impact} shows no clear
correlation.  In three of the confirmed DLA galaxies optical emission
lines have velocity differences between --200 and 30~km~s$^{-1}$
relative to \lya\ \citep{weatherley05}.  Some candidates have larger
offsets, possibly affected more strongly by resonant scattering.

\subsection{\ion{H}{i} extension}
\label{sect:hiz_imp}
The average impact parameter of 16 kpc derived for the candidates is
larger than that expected by numerical simulations which favor impact
parameters of $b=3$ kpc for DLA galaxies, and have fewer than 25\%
with $b>10$~kpc at all redshifts \citep{haehnelt00,okoshi04,hou05}.
Larger DLA galaxy sizes of 10--15 kpc at $2<z<4$ are inferred by other
simulations \citep{gardner01}. A possibility for the difference
between observations and simulations is that the simulations assume a
single disk scenario, while DLA galaxies could exist in groups
\citep{hou05}. The real absorbing galaxy could be fainter and lie
closer to the QSO line of sight than the detected candidate galaxy.

An anticorrelation between \nhi\ and the distance to the nearest
galaxy is found in simulations \citep{gardner01}, but no analysis of
this effect for observed DLA galaxies has been attempted.  A trend for
larger column density absorbers at smaller impact parameters was
observed in a sample of DLA galaxies at $z<1$ \citep{rao03}.  At lower
column densities in the \lya\ forest such an anticorrelation has been
shown to exist \citep{chen98,chen01}.  Observations of the galaxies
giving rise to \ion{Mg}{ii} absorption lines showed an anticorrelation
between the impact parameters and column densities of both
\ion{Mg}{ii} and \ion{H}{i} \citep{churchill00}, but recent
observations of a larger sample have indicated that the correlation is
not always present \citep{churchill05}.

We here investigate whether the candidates show a similar
anticorrelation using the impact parameters for the candidates as a
proxy for the sizes of neutral gas envelopes around proto galaxies.
Assuming such a correlation is necessarily a rough approximation
because large morphological differences between individual systems are
expected \citep{rao03,chen03}.  Specifically, the possible presence of
sub-clumps is neglected.

\subsubsection{DLA galaxy sizes and luminosities}
\begin{figure}
\centering
\resizebox{\hsize}{!}{\includegraphics[bb= 0 0 566 425,clip]{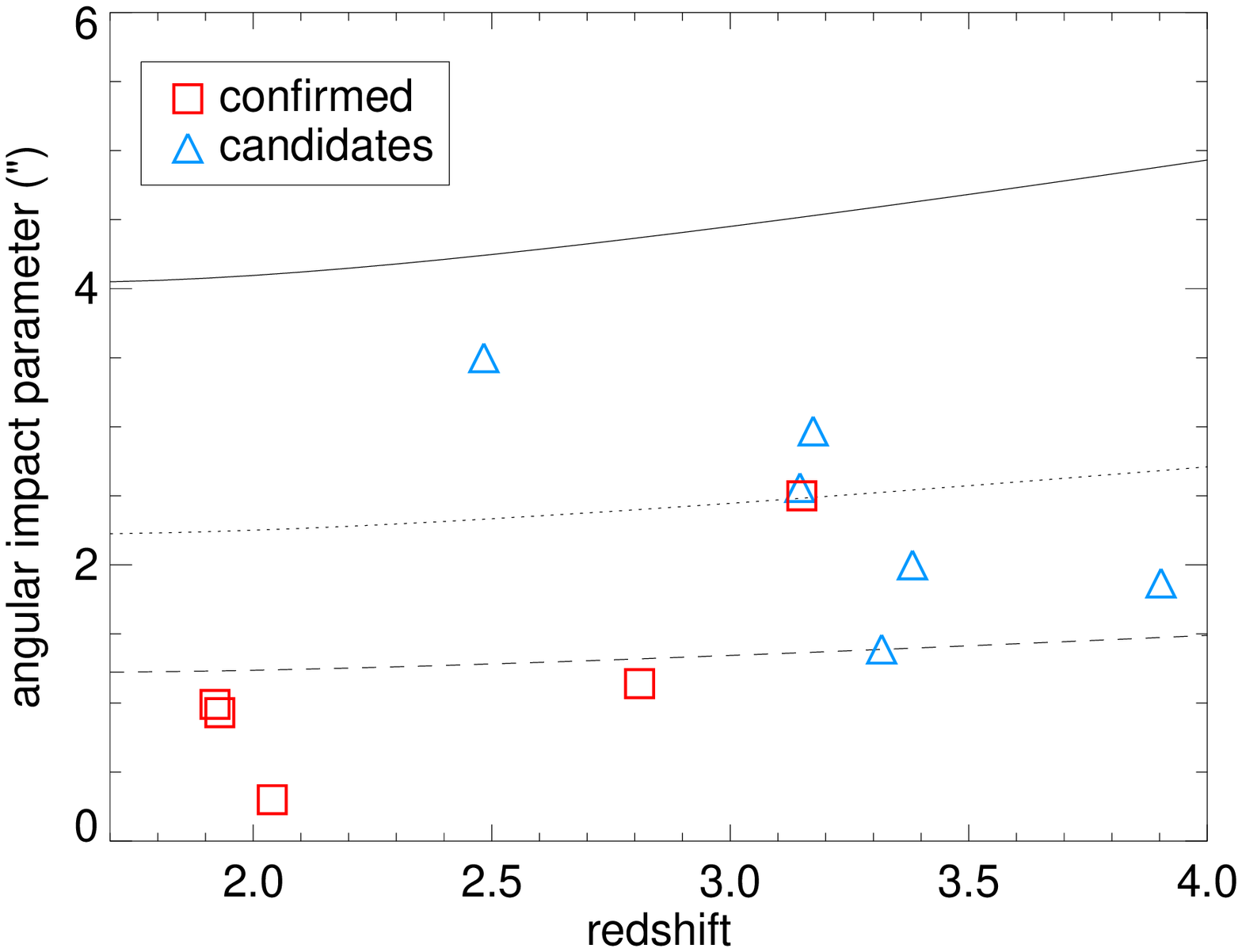}}
\caption{Angular impact parameter vs. redshift. The solid line
  corresponds to the size-luminosity relation for an $L_B^*$ galaxy
  \citep{chen03}, while the dotted and dashed lines correspond to
  galaxies with $B$ band luminosities $L_B=10\%L_B^*$ and
  $L_B=1\%L_B^*$, respectively. Squares are objects from the
  literature and triangles the candidates from this survey.}
\label{fig:hiz_ang}
\end{figure}

To analyse the sizes of DLA galaxies at $z<1$, \citet{chen03} describe
the extension of the neutral gas cloud associated with DLA galaxies as
\begin{equation}
 \frac{R}{R^*}=\Big(\frac{L_B}{L_B^*}\Big)^t ,
\label{eq:hiz_scale_r}
\end{equation}
assuming that DLA galaxies follow a Holmberg relation between galaxy
sizes and luminosities.  Their fit to the observed DLA galaxies gives
$R^*=30h^{-1}$~kpc, $t=0.26_{-0.06}^{+0.24}$, where $L_B^*$
corresponds to a galaxy with $M_B^*=-19.6$. Based on the morphologies
and impact parameters \citet{chen05} argued that dwarf galaxies alone
cannot represent the DLA galaxy population at $z<1$.  In the case that
the DLA galaxy population evolves from low luminosity objects at high
redshifts to higher luminosity at lower redshifts, this will affect
the expected impact parameters. In Fig.~\ref{fig:hiz_ang} the impact
parameters of candidates and confirmed objects are shown as a function
of their redshifts. Overlayed on this figure are curves for the
size-luminosity relation derived for low redshift DLA galaxies. If DLA
galaxies at high redshift follow the low redshift scaling relation,
they comprise a mix of galaxy luminosities.

\subsubsection{Powerlaw profiles}
\begin{figure}
\centering
\resizebox{\hsize}{!}{\includegraphics[bb= 0 0 566 425,clip]{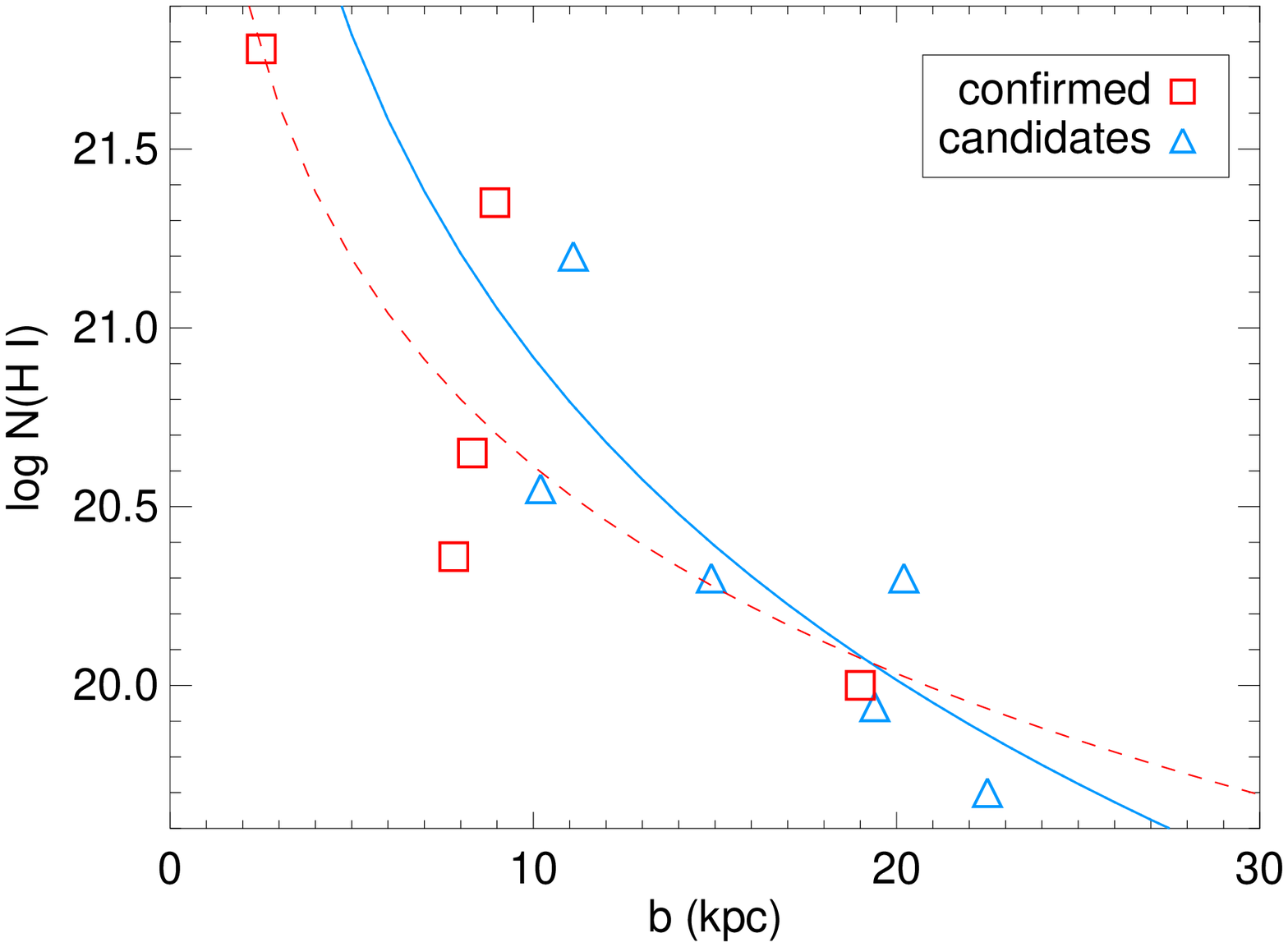}}
\caption{ Column density of neutral hydrogen as a function of impact
  parameters for the candidates and previously confirmed objects. The
  symbols are similar to the previous figures.  The solid and dashed
  lines are fits to the power-law relation \(b/b^*=(N/N^*)^{\beta}\)
  for the candidates and confirmed objects, respectively.  }
\label{fig:hiz_imp_nhi}
\end{figure}

Using similar arguments as above one could expect that there is a
relation between the impact parameter $b$, and the column density
measured for the DLA.  Fig.~\ref{fig:hiz_imp_nhi} shows the
\nhi\ measured for the DLA as a function of the impact parameters in
kpc. We assume a similar scaling relation as in
Eq.~(\ref{eq:hiz_scale_r}) for the impact parameter and \nhi, i.e.
\begin{equation}
\frac{b}{b^*}=\Big(\frac{\nhi}{\nhi^*}\Big)^{\beta}
\label{eq:hiz_b_nhi}
\end{equation}

We set $\log$~\nhi$^*$~=~20.3 and the error on the measured impact
parameter is given by the fibre size of 0\farcs5, which corresponds to
$\sim$4 kpc. A fit of the observed impact parameters for the
candidates gives $b^*=15.9\pm1.4$ kpc, and $\beta=-0.23\pm0.08$ as
shown by the solid line, while for the confirmed objects, we find
$b^*=12.0\pm3.7$ kpc and $\beta=-0.36\pm0.14$ as represented by the
dashed line. The fits to the candidates and confirmed objects are
consistent within 1$\sigma$ uncertainties.

\subsubsection{Exponential profiles}
\label{sect:hiz_exp}
\begin{figure}
\centering
\resizebox{\hsize}{!}{\includegraphics[bb= 0 0 566 425,clip]{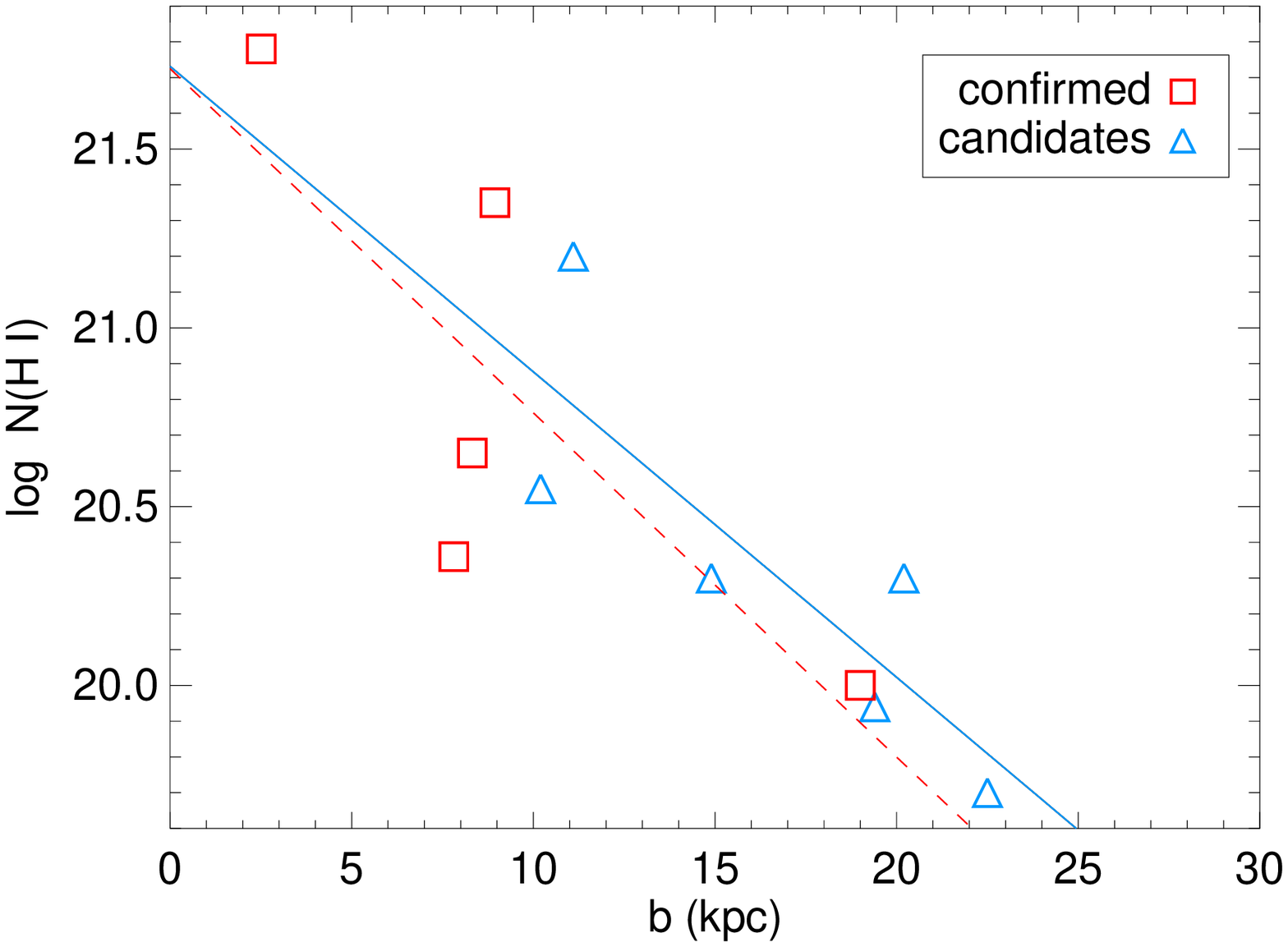}}
\caption{Column density of neutral hydrogen as a function of impact
  parameter for the candidates and previously confirmed objects.  The
  solid and dashed lines show the fit to the exponential relation \(
  \nhi = \nhi_0 \exp (-b/h) \) for the candidates and confirmed
  objects, respectively.}
\label{fig:hiz_exp_profile}
\end{figure}

Radio observations of the 21~cm emission from \ion{H}{i} disks in the
local Universe have shown that an exponential profile is a poor
representation in the central part of disk galaxies, where the 21 cm
flux density either stays constant or decreases towards the centre
\citep[e.g.][]{verheijen01}. However, at optical wavelengths disk
galaxies are well represented by exponential profiles. Here we fit the
impact parameter distribution by the relation
\begin{equation}
\nhi= \nhi_0 \exp(-b/h)
\end{equation}
where $h$ is the scale length, $\nhi_0$ the central column density of
a simple exponential disk, and \nhi\ is the column density measured
for the DLAs. The resulting fit to all the candidates is shown by the
solid line in Fig.~\ref{fig:hiz_exp_profile}, which has $\log
\nhi_0=21.7\pm1.1$~cm$^{-2}$ and $h=5.1^{+2.5}_{-1.3}$ kpc. This
result is similar within the uncertainties to a fit to the confirmed
objects only ($\log N_0=21.7\pm1.1$~cm$^{-2}$ and
$h=4.5^{+3.6}_{-1.4}$ kpc).

Local disk galaxies have optical scale lengths ranging from $\sim$2
kpc to $\sim$6 kpc, and observations of the \ion{H}{i} profile in low
surface brightness galaxies have indicated scale lengths $>10$ kpc
\citep{matthews01}. In contrast, higher redshift spiral galaxies in
the HST Ultra Deep field have smaller optical scale lengths of
1.5--3~kpc \citep{elmegreen05}, possibly biased towards smaller values
due to the easier detection of high surface brightness, high star
formation rate regions. The question is how extended the gaseous
envelopes are around these young galaxies. The impact parameters of
the candidates suggest that high redshift DLAs reside far from the
host galaxy, if not in a regular proto-galactic disk \citep{wolfe86},
then in a region of the same physical scale, possibly in merging
clumps of gas surrounding the actual proto galaxy. In this picture,
DLAs can be found far from the center of the parent galaxy.

The two objects that were originally discarded as candidates due to
their large impact parameters ($>30$ kpc) do not follow the relations
for either the exponential or power-law profiles. Including them would
make the scatter around the fit substantial.

\subsection{Metallicity effects on \lya\ emission}
Using a space based imaging survey and follow up long-slit
spectroscopic observations, \citet{moller04} found indications for a
positive metallicity--\lya\ luminosity relation, such that
\lya\ emission was preferentially observed in higher metallicity
systems. They argued that this positive correlation could over-power
the negative dust--\lya\ luminosity effects which are expected to be
strong in high metallicity environments \citep{charlot93}. Studies of
\lya\ emission from nearby star-forming galaxies have not revealed any
correlations between metallicity and \lya\ emission strength
\citep{keel04}. Differences in the velocity-metallicity relation
between high and low redshift DLAs could be explained by higher
redshift DLAs residing in lower mass galaxies \citep{ledoux06}, that
have fainter \lya\ emission.

\begin{figure}
\centering \resizebox{\hsize}{!}{\includegraphics{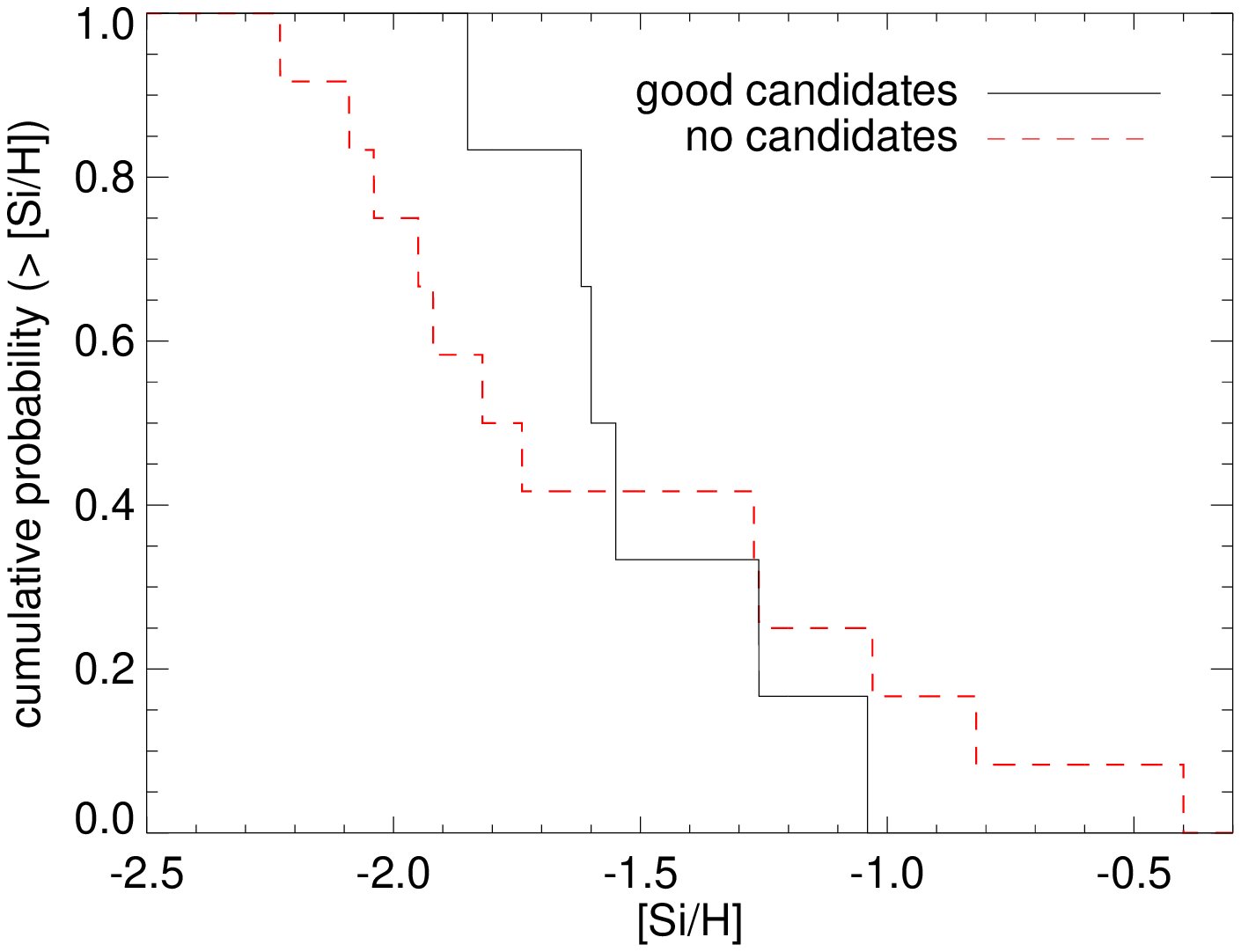}}
\caption{Cumulative distribution of Si metallicities of the six good
  emission line candidates compared to the remaining part of the
  sample. The probability that the two distributions are similar is
  38\% estimated from a Kolmogorov-Smirnov test.}
\label{fig:hiz_cumdist_si}
\end{figure}

In this context we investigate the distribution of metallicities for
the emission line candidates in comparison to the total sample.  We
compare the cumulative distributions of metallicities ([Si/H]) for the
DLAs that have candidate \lya\ detections with the metallicities of
the remaining objects that have either no \lya\ candidates or rejected
candidates. The distributions in Fig.~\ref{fig:hiz_cumdist_si} show
the fraction of DLAs with metallicities larger than a given value.
Table~\ref{tab:list_obj} has several lower limits on [Si/H] and
Fig.~\ref{fig:hiz_cumdist_si} treats the limits as actual detections.

A two sided Kolmogorov Smirnov (KS) test gives a probability of 38\%
that the two samples have the same underlying distributions.  A
similar analysis for [Fe/H] gives the same result. Hence, none of the
tests give clear statistical evidence for a difference between the two
populations.  Only a small number of DLAs are included in this survey.
For the two samples with $N_1$ and $N_2$ being the number of objects
in each sample respectively we find $N_1N_2/(N_1+N_2) =4$. For the KS
test to be statistically valid a value larger than 4 is required
\citep{press92}, hence a few more objects are needed to make the test
more statistically significant.

\subsection{Metallicity gradients}
\label{sect:hiz_met}
In local galaxies the metallicities of \ion{H}{ii} regions are shown
to decrease with increasing radial distance in the disk
\citep{zaritsky94}.  If DLAs arise in disks lower metallicities are
expected at larger linear separations between the QSO line of sight
and the galaxy centers. A comparison of absorption metallicities for 3
DLAs at $z<0.6$ with abundances derived from strong emission line
diagnostics from the galaxy spectra revealed that gradients are likely
present at at level --0.041$\pm$0.012~dex~kpc$^{-1}$ \citep{chen05}.
Uncertainties in the gradients arise due to an unknown correction for
dust depletion and also the inclination of the galaxy plays an
important role due to projection effects.  Metallicity differences due
to differential depletion within a singe DLA galaxy can be strong as
demonstrated by observations of a lensed QSO \citep{lopez05}.

Observations of nebular emission lines from seven galaxies at
$2.0<z<2.5$ indicate an average solar metallicity \citep{shapley04},
with evidence for the presence of metallicity gradients
\citep{foerster-schreiber06}.  If metallicity gradients exist for high
redshift DLA galaxies, we would expect to see a tendency for higher
metallicities for the DLA galaxies detected at smaller impact
parameters.  DLA galaxies are on the average fainter than LBGs
detected in flux limited surveys \citep{moller02}.  Combining this
with a high redshift luminosity-metallicity relation can imply low DLA
metallicities without involving metallicity gradients.

\begin{figure}
\centering
\resizebox{\hsize}{!}{\includegraphics{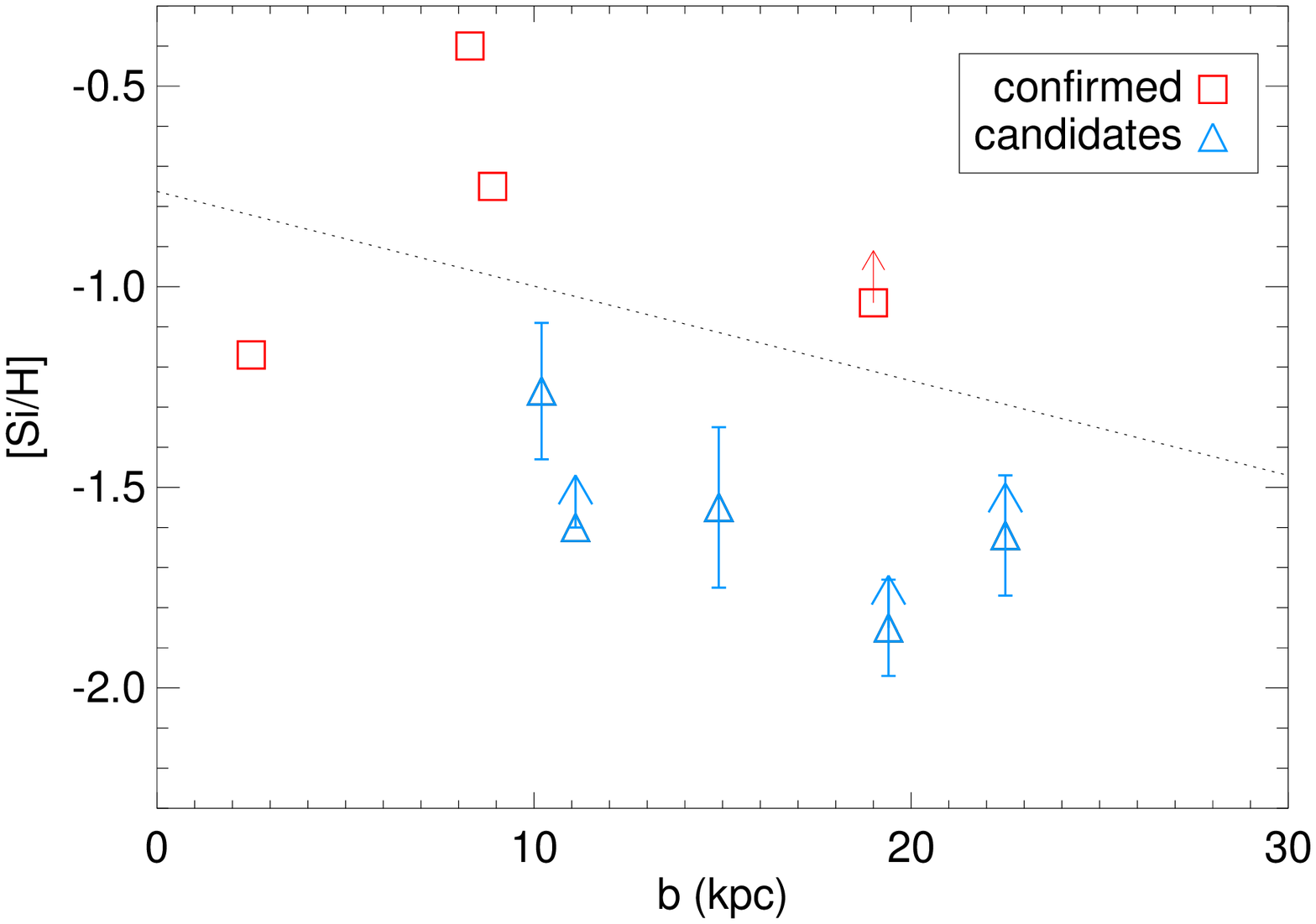}}
\caption{DLA metallicities as a function of the impact parameters,
  where symbols shapes have similar meanings as in the previous
  figures. The dotted line shows a fit to all the objects excluding
  the limits. This line has a gradient of
  $-0.024~\pm~0.015$~dex~kpc$^{-1}$. Data for the confirmed objects
  are either [Si/H] of [Zn/H] taken from \citet{moller04}.}
\label{fig:hiz_metal_grad}
\end{figure}

Fig.~\ref{fig:hiz_metal_grad} shows the DLA metallicities as a
function of the impact parameters. The line shows a fit to all objects
ignoring those with lower limits on [Si/H].  This gradient is
--0.024$\pm$0.015, i.e. is consistent with zero at the 2$\sigma$
level. The large scatter in the plot could be real and unrelated to
gradients. Different DLAs exhibit a large range of star formation
histories \citep{erni06,herbert-fort06}, which makes it unreasonable
to expect a smooth relation between metallicities and impact
parameters for a sample of DLAs. Clearly more data are needed to
determine the reality of any relation.

\section{Conclusions}
\label{sect:hiz_conc}

We have presented an integral field spectroscopic survey of 9 high
redshift QSOs, which have a total of 14 DLA systems and 8 sub-DLA
systems.  We detect eight good candidates for \lya\ emission lines
from DLA and sub-DLA galaxies. Two of these are found at impact
parameters larger than 30 kpc, and are not likely associated directly
with the absorbing galaxy, but could be associated with galaxy groups
in which the real absorbing galaxy resides.  All candidates are
detected at a statistically significant level in reconstructed
narrow-band images as well as in the co-added one-dimensional spectra.
Further observations will be useful to independently confirm the
candidates at an even higher signal to noise ratio.  We compare the
properties inferred from the IFS data with those for previously
spectroscopically confirmed \lya\ emission lines from DLA galaxies
reported in the literature.  We find that line luminosities are
similar to those of previously confirmed objects, that the average
impact parameters are larger by a factor of $\sim$2, and that some
candidates have larger velocity offsets between the \lya\ emission
line and the systemic redshift of the DLA system.

We analyse the distribution of DLA column densities as a function of
impact parameters. Assuming that the average DLA galaxy is similar to
a disk galaxy with an exponential profile, we show that it has a scale
length of 5 kpc. Such a scale length is similar to disk scale lengths
found for local spiral galaxies.  This could imply that DLAs belong to
large disks even at high redshifts as originally suggested by
\citet{wolfe86}. However, it is probably too simplistic to expect that
high redshift DLAs reside in regular disks with similar structure to
large local disks.  DLA systems are generally not associated with
luminous galaxies \citep[e.g.][]{colbert02,moller02}.  The large
impact parameters found for the candidates indicate that the
distribution of \ion{H}{i} clouds in DLA galaxies extends
significantly beyond the optical sizes of fainter dwarf galaxies.

Furthermore, \citet{wolfe06} showed that high redshift DLAs do not
reside in extended disks that follow the local Schmidt-Kennicutt law
for star formation. The IFS results presented here suggest that the
\lya\ emission is generally not extended, and that star formation
takes place at a distance of several kpc from the DLA. Hence, we may
speculate that DLAs arise in the outskirts of proto-galaxies, for
example in clouds of neutral gas around LBGs. In this case one would
expect a significant scatter in the relation between the impact
parameter and column density of the DLA since neutral clouds could be
distributed irregularly around the galaxy. Contrary to expectations,
the objects show a small scatter around the relations in
Figs.~\ref{fig:hiz_imp_nhi} and ~\ref{fig:hiz_exp_profile}.
Regardless of the distribution of neutral gas in DLA galaxies we
conclude that there is a tendency to find a lower column density DLA
with increasing impact parameter. Extending the investigation to
include DLAs and sub-DLAs with \nhi$> 10^{19.6}$~cm$^{-2}$ this
tendency emerges for both the candidates and the confirmed objects.

The velocity offsets between the \lya\ emission lines and the systemic
redshifts of the DLAs are larger for half of the candidates compared
to the confirmed objects. This could indicate an origin in groups of
galaxies, where the DLA resides in a less luminous component than the
galaxy detected in \lya.  To determine whether resonant scattering
affects the candidate \lya\ lines more strongly and gives rise to
larger velocity offsets than for the confirmed objects, observations
of the corresponding optical emission lines which are shifted to the
near-IR are required \citep[e.g.][]{weatherley05}.  Optical emission
lines furthermore have the advantage of being less affected by dust
absorption and therefore are better for estimating the star-formation
rates. Alternatively non-resonance UV lines such as \ion{C}{iv} could
be studied.

This survey was carried out with IFS on a 4-m class telescope, and the
signals were generally near the detection limit. To verify this IFS
method, it is necessary to get independent, higher signal-to-noise
ratio spectra with a larger aperture telescope to confirm the
existence of the emission lines.

\begin{acknowledgements}
  This study was supported by the German Verbundforschung associated
  with the ULTROS project, grant no. 05AE2BAA/4. S.F.~S\'anchez
  acknowledges the support from the Euro3D Research Training Network,
  grant no. HPRN-CT2002-00305. K. Jahnke acknowledges support from DLR
  project No.\ 50~OR~0404. We thank the referee for useful suggestions
  that clarified the paper.
\end{acknowledgements}

\bibliography{ms6410}
\end{document}